\documentclass[pra,nofootinbib,preprintnumbers,tightenlines,twocolumn,superscriptaddress]{revtex4-1}
\usepackage{braket}
\usepackage{array}
\usepackage{dsfont}
\usepackage{amsmath}
\usepackage{pifont}

\usepackage{amssymb}

\usepackage{amsfonts}
\usepackage{mathtools}
\usepackage{graphicx}
\usepackage{dcolumn}
\usepackage{bm}
\usepackage{multirow}

\usepackage{leftidx}
\usepackage{color}
\usepackage{mathtools}
\usepackage{MnSymbol}
\usepackage[mathscr]{eucal}
\usepackage[german,english]{babel}
\usepackage[capitalize]{cleveref}

\usepackage[utf8]{inputenc}
\usepackage{natbib}
\usepackage{comment}
\newcommand{\abs}[1]{\left| #1 \right|}
\begin{document}

\title{Few-body correlations in two-dimensional Bose\\ and Fermi ultracold mixtures}

\author{G. Bougas}
\email{gbougas@physnet.uni-hamburg.de}
\affiliation{Center for Optical Quantum Technologies, Department of Physics, University of Hamburg, Luruper Chaussee 149, 22761 Hamburg Germany }

\author{S. I. Mistakidis}
\affiliation{Center for Optical Quantum Technologies, Department of Physics, University of Hamburg, Luruper Chaussee 149, 22761 Hamburg Germany }

\author{P. Giannakeas}
\affiliation{Max-Planck-Institut f\"ur Physik komplexer Systeme, N\"othnitzer Str.\ 38, D-01187 Dresden, Germany }

\author{P. Schmelcher}
\affiliation{Center for Optical Quantum Technologies, Department of Physics, University of Hamburg, Luruper Chaussee 149, 22761 Hamburg Germany }
\affiliation{The Hamburg Centre for Ultrafast Imaging, University of Hamburg, Luruper Chaussee 149, 22761 Hamburg, Germany}

\begin{abstract}
Few-body correlations emerging in two-dimensional harmonically trapped mixtures, 
are comprehensively investigated. The presence of the trap leads to the formation of atom-dimer and trap states, in addition to trimers. The Tan's contacts of these eigenstates are studied for varying interspecies scattering lengths and mass ratio, while corresponding analytical insights are provided within the adiabatic hyperspherical formalism. The two- and three-body correlations of trimer states are substantially enhanced compared to the other eigenstates. 
The two-body contact of the atom-dimer and trap states features an upper bound regardless of the statistics, treated semi-classically and having 
an analytical prediction  
in the limit of large scattering lengths. Such an upper bound is 
absent in the three-body contact. Interestingly, by tuning the interspecies scattering length the contacts oscillate as the atom-dimer and trap states change character through the existent avoided-crossings in the energy spectra. For thermal gases, a gradual suppression of the involved two- and three-body correlations is evinced manifesting the impact of thermal effects. Moreover, spatial configurations of the distinct eigenstates ranging from localized structures to angular anisotropic patterns are captured. Our results provide valuable insights into the inherent correlation mechanisms of few-body mixtures which can be implemented in recent ultracold atom experiments and will be especially useful for probing the crossover from few- to many-atom systems.

\end{abstract}
\maketitle

\section{Introduction}

The advent of optical tweezers corroborates the experimental realization of few-body ultracold atom settings~\cite{Bayha_shell_2020,Holten_Crystalls_2020,blume_few-body_2012} in a controllable manner even at the level of two~\cite{Xu_interaction_decay_2015,Sompet_microtrap_2013,Anderegg_tweezer_2019,Guan_density_2019}, and three \cite{Reynolds_triads_2020} atoms. Moreover, advances in the relevant trapping techniques provide an exquisite variability of such systems, e.g. in terms of reduced dimensionality \cite{Bayha_shell_2020,Holten_Crystalls_2020} or tunable atomic interactions through Feshbach \cite{chin_feshbach_2010} and confinement induced resonances \cite{olshanii_atomic_1998,bergeman_atom-atom_2003,haller_confinement-induced_2010,giannakeas_coupled_2012}. As such, strongly correlated few-body systems are nowadays accessible with a high fidelity in a prosaic way.

Three-particle systems in two-dimensions (2D) are of particular interest given that they yield insights into the stability properties of 2D gases in terms of their inherent three-body recombination processes \cite{Incao_recomb2D_2015,Helfrich_resonant_2011,Pricoupenko_stability_2007}, and are viewed as the fundamental building-blocks for understanding the crossover from few- to many atom systems \cite{liu_correlated2D_2010,daily_hyperspherical_2015,greene_clusters_2017,gharashi_three_2012}. 
Also, they constitute the minimal settings containing both two- and three-body correlations whose characteristics are essential for engineering many-body processes \cite{daily_hyperspherical_2015,Kirk_three-body_2017,Bermudez_MB_2009}. 
The reduced dimensionality plays a crucial role on the impact of correlations, namely they are more prominent in lower compared to three-dimensions (3D)~\cite{Lang_correlations_2017,Lindgren_fermionization_2014,zinner_fractional_2014}. 
Conventionally, correlation effects manifest in the asymptotic expansion of the momentum distribution of the one-body reduced density~\cite{Bellotti_contacts_2014,bellotti_dimeffects_2013} and are consecutively captured by the so-called two- and three-body contacts.
The latter are experimentally probed via radio-frequency spectroscopy, time-of-flight expansion, and subsequent measurement of the structure factor with the aid of Bragg spectroscopy \cite{Fletcher_contacts_2017,wild_measurement_2012,sagi_measurement_2012,stewart_contact_2010,Kuhnle_universal_2010,Hoinka_precise_2013}. Their investigation sheds light into the microscopic properties of the system, especially the formation of two- \cite{sykes_quenching_2014,Corson_bound_2015,Bougas_analytical_2019,Bougas_stationary_2020} and three-body bound states (trimers) \cite{bellotti_dimeffects_2013,sykes_quenching_2014}. Importantly, the two-body contact satisfies universal relations regarding the energy, the two-body loss rate, and the radio-frequency spectra that hold regardless of the statistics and the dimensionality, in few- as well as in many-body settings \cite{castin_general_2012,valiente_universal_2011,Valiente_universal_2012,Werner_generalF_2012}. 

Three-body correlations on the other hand, captured by the three-body contact, strongly depend on the dimensionality of the system \cite{bellotti_dimeffects_2013,Bellotti_contacts_2014}. This behavior is attributed to the presence of the Efimov effect in 3D, which significantly affects the energy spectra of three-body systems \cite{bellotti_dimeffects_2013,castin_single-particle_2011}, and in particular the trimer states \cite{greene_clusters_2017}. Interestingly, the experimental observation of the three-body contact in 2D settings remains, to the best of our knowledge, yet elusive. The important role of correlations in reduced dimensions however renders its study of immense interest, especially in trapped three-particle systems. This is further corroborated by the investigation of the two-body contact, in strongly interacting harmonically trapped two-component Fermi gases \cite{frolich_liquid_2012,bertaina_bcs_2011}, and more recently in a two-component bosonic gas confined in a 2D box potential \cite{zou_planar_Cont_2020}, revealing enhanced two-body correlations in the BEC-BCS crossover.

Particularly, 2D binary set-ups consisting of two identical bosonic or fermionic atoms interacting with a third distinguishable one are known to possess a plethora of trimer states in terms of the 2D scattering lengths among the identical particles (intraspecies) and the two different atoms (interspecies), as well as the mass ratio \cite{bellotti_mass-imbalanced_2013,bellotti_scaling_2011}. This holds in spite of the absence of the Efimov effect \cite{lim_fonseca-redish-shanley_1980,bruch_binding_1979}, which is usually manifested as the appearance of an infinite progression of trimer states. Another crucial property is that when the identical bosons or fermions become heavier than the third particle, ancillary trimer states are created \cite{bellotti_mass-imbalanced_2013,pricoupenko_planar_2010}.
Generally, studies in 3D have shown that the mass ratio can drastically affect the properties of the three-body complexes allowing, for example, more favorable experimental conditions to observe multiple successive Efimov states \cite{tung_geometric_2014, pires_PRL_2014}, or resonant effects that are absent on equal mass three-body collisions \cite{ulmanisHeteronuclearEfimovScenario2016,johansen_NP_2017,wackerUniversalThreeBodyPhysics2016, giannakeas_PRL_2018,mikkelsenThreebodyRecombinationTwocomponent2015,petrov_PRA_2015}.
Apart from that, the confinement of three-body systems in a 2D harmonic oscillator yields the presence of additional eigenstates in the energy spectrum aside from trimer ones. These consist of a dimer interacting with another trapped atom \cite{liu_correlated2D_2010}, as well as trap states characterizing three weakly interacting atoms confined in a harmonic potential.

Focusing on the problem of 2D three-particle binary settings, we unveil their emergent few-body correlation characteristics for a wide range of their intrinsic parameters such as the interspecies scattering lengths and the mass ratio, as well as for different particle statistics.
An emphasis is placed on the impact of the above-described additional eigenstates originating from the presence of the trap which has not been studied so far \cite{Bellotti_contacts_2014}. 
Exploiting the utility of the adiabatic hyperspherical formalism \cite{greene_clusters_2017,naidon_efimov_2017,dincao_few-body_2017}, the corresponding few-body correlation measures are constructed and their behavior, as captured by the two- and three-body contacts, is systematically investigated  over a vast parameter space which is comprised by the particles' statistics, the 2D scattering lengths and the mass ratio.

In particular, for trimer states we observe the same overall behavior of the corresponding two- and three-body correlations, as was shown in previous studies \cite{Bellotti_contacts_2014}.
However, we explicate that despite the statistics, both the two- and three-body contacts of atom-dimer and trap states display an oscillatory pattern for varying scattering length. 
This behavior is attributed to the existent avoided-crossings between these two eigenstates in the energy spectra.
By considering thermal gases, the amplitude of these oscillations decreases for larger temperatures, a phenomenon that holds equally for the magnitude of two- and three-body correlations. 
Interestingly, the atom-dimer states provide an upper bound for the two-body contact of all non-trimer states and a semi-analytical prediction is derived within the JWKB method, regardless of the particle exchange symmetry. Such a bound is absent in the case of three-body correlations. Binary systems with bosonic majority species exhibit overall an increased degree of correlations, due to the existence of three-body ones, being absent in their fermionic counterpart. Moreover, the spatial configuration of the eigenstates is demonstrated via the experimentally accessible one-body reduced density in position space, an observable largely unexplored in 2D three-body systems \cite{sandoval_radii2D_2016}. 

This work is arranged as follows. In Sec. \ref{Sec:Hamilt}, the Hamiltonian of the considered mixtures is introduced within the hyperspherical formalism whose main aspects are presented in detail. In Sec. \ref{Sec:Spectra} we review the behavior of the adiabatic potential curves stemming from the hyperangular problem and the underlying energy spectra for different scattering lengths and mass ratios. 
Subsequently, the susceptibility of the contacts is unraveled with respect to the scattering lengths [Sec. \ref{Sec:Cont2}] and the mass ratios [Sec. \ref{3b_contact}]. Furthermore, the spatial configuration of the binary 2D three-body systems is revealed via the reduced one-body density in Sec. \ref{Sec:dens}. 
We conclude and discuss future perspectives in Sec. \ref{Sec:conclusions}. 
Appendix \ref{Ap:Hyperangles} discusses the boundary condition of two colliding particles within the hyperspherical formalism while Appendix \ref{Ap:Asymptotic} elaborates on the derivation of the reduced one-body density and its asymptotic expansion in momentum space. Finally, in Appendix \ref{Ap:Upper_bound} an analytic bound is established for the two-body contact of atom-dimer states for all binary mixtures.

\section{Hamiltonian and hyperspherical framework}  \label{Sec:Hamilt}

In the following, we focus on the three-body collisions of harmonically trapped binary mixtures in 2D.
The three-body collisional complex mainly consists of two identical particles of either bosonic or fermionic symmetry and a third distinguishable one where their pairwise interactions are modeled via $\delta-$function pseudopotentials. 
This setup constitutes a straightforward generalisation of the analytically tractable trapped two-body problem ~\cite{busch_two_1998,Budewig_Quench_2019,Farrell_universality_2009,Bougas_analytical_2019}.
These particular considerations permit us to investigate the dependence of two-/three-body correlations on the scattering lengths, the mass ratio of the particles as well as the impact of particle symmetry.
In view of the broad parameter space, three-body collisions in 2D are best treated in the theoretical framework of the adiabatic hyperspherical approach. 
One particular aspect of this method is that the particle symmetry can be postimposed.
Therefore, following Refs. ~\cite{rittenhouse_hyperspherical_2016,rittenhouse_greens_2010} the general scope of the hyperspherical approach is presented below whereas the particle symmetry is imposed at the end of this section.

In the laboratory (lab) frame the Hamiltonian of three-particles of mass $m_i$ ($i=1,2,3$) in a 2D isotropic trap of frequency $\omega$ reads:

	  \begin{equation}
	     \mathcal{H}=\sum_{i=1}^{3} \left( -\frac{\hbar^2}{2m_i} \nabla_i^2 +\frac{1}{2}m_i \omega^2 \boldsymbol{r}_i^2 \right)+\sum_{i<j} V_{ij}(\boldsymbol{r}_i-\boldsymbol{r}_j),
	     \label{Eq:hamilt_lab}
	  \end{equation}
	  where $\boldsymbol{r}_i$ is the 2D position vector of the $i$-th particle with mass $m_i$. The regularized pseudopotential describing pairwise $s$-wave interactions in 2D~\cite{olshanii_pseudopot2D_2001} 
	  is given by
	  \begin{equation}
	    V_{ij}(\boldsymbol{r}_{ij})= -\frac{\pi \hbar^2 \delta^{(2)}(\boldsymbol{r}_{ij})}{\mu_{ij}\ln (A \lambda a^{(k)})}\left[1-\ln (A\lambda r_{ij})r_{ij}\frac{\partial}{\partial r_{ij}}\right],
	    \label{Eq:Pseudpot}
	  \end{equation}
where $\boldsymbol{r}_{ij}=\boldsymbol{r}_i-\boldsymbol{r}_j$ and the reduced two-body mass is given by $ \mu_{ij}=\frac{m_im_j}{m_i+m_j}$. The constant $A$ is $A=e^{\gamma}/2$ where $\gamma\approx0.577$ is the Euler-Mascheroni constant. Also, $\lambda$ serves as an ultraviolet-cutoff for the zero-range pseudopotential and provides an upper bound in momentum, which, however, does not impact any observable \cite{olshanii_pseudopot2D_2001}.
$a^{(k)}\equiv a_{ij}$ refers to the 2D scattering length between the particles $i$ and $j$, labeled in the odd-man-out notation \cite{rittenhouse_greens_2010}. Note that $a^{(k)}$ is related to the 3D scattering length, $a_{3D}^{(k)}$, via the expression $a^{(k)}=2e^{-\gamma}\sqrt{\pi/0.915}l_0^{(k)}\exp\{-\sqrt{\pi/2} l_0^{(k)}/a_{3D}^{(k)}\}$ \cite{petrov_interatomic_2001}. Here, $l_0^{(k)}=\sqrt{\hbar/(\mu_{ij} \omega_z)}$ denotes the harmonic oscillator length in the $z$-direction perpendicular to the 2D plane. Experimentally, $a_{3D}^{(k)}$ can be flexibly tuned by means of Feshbach resonances \cite{chin_feshbach_2010}. 

\begin{figure}[t]
    \centering
    \includegraphics[width=0.5 \textwidth]{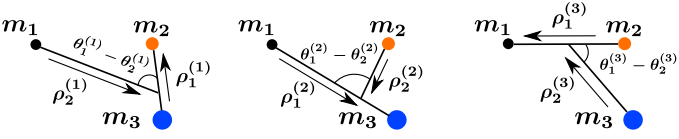}
    \caption{Sketch of the three possible sets of Jacobi vectors regarding three distinguishable atoms with masses $m_1$, $m_2$ and $m_3$. $\theta_{1,2}^{(k)}$ are the polar angles corresponding to the $\boldsymbol{\rho}_{1,2}^{(k)}$ Jacobi vectors [Eq.~(\ref{Eq:Jacobi})] which connect the atoms $i$ and $j$, $\boldsymbol{\rho}_1^{(k)}$, and their center-of-mass with the third atom $k$, $\boldsymbol{\rho}_2^{(k)}$.}
    \label{Fig:Jacobi}
\end{figure}
The number of degrees of freedom in the Hamiltonian of \cref{Eq:hamilt_lab} can be reduced by changing from the lab- to the body-frame  of reference. 
This permits us to separate $\mathcal{H}$ into center-of-mass and relative Hamiltonian contributions.
This can be achieved by transforming the lab coordinates $\mathbf{r}_i$, with $i=1,2,3$, into a set of three equivalent mass-scaled Jacobi vectors \cite{greene_clusters_2017,whitten_symmetric_1968}. Namely
\begin{eqnarray}
  & &\boldsymbol{r}_{\textrm{CM}}=\frac{m_1\boldsymbol{r}_1+m_2\boldsymbol{r}_2+m_3\boldsymbol{r}_3}{m_1+m_2+m_3}, \label{Eq:CM} \\
 & &\boldsymbol{\rho}^{(k)}_1=\frac{\boldsymbol{r}_i-\boldsymbol{r}_j}{d_k}, \\
 & & \boldsymbol{\rho}^{(k)}_2=d_k\left( \frac{m_i \boldsymbol{r}_i+m_j\boldsymbol{r}_j}{m_i+m_j}-\boldsymbol{r}_k\right),
 \label{Eq:Jacobi}
\end{eqnarray}
where $d_k^2=\frac{m_k(m_i+m_j)}{\mu(m_1+m_2+m_3)}$ and $\mu=\sqrt{\frac{m_1m_2m_3}{m_1+m_2+m_3}}$ is the three-body reduced mass.
The superscript $(k=1,2,3)$ labels the three sets of the relevant Jacobi vectors. The first vector,  $\boldsymbol{\rho}_1^{(k)}$, links the particle pair $i-j$ whereas $\boldsymbol{\rho}_2^{(k)}$ relates the $k$-th particle with the center-of-mass of the pair~\cite{rittenhouse_greens_2010}, see also Fig. \ref{Fig:Jacobi}. 

The separability of the center-of-mass and relative degrees of freedom permits us to consider that the center-of-mass part of the  total wave function, namely $\Psi_{\textrm{CM}}(\boldsymbol{r}_{CM})$, resides in its ground state, i.e. $\Psi_{\textrm{CM}}(\boldsymbol{r}_{CM})=\sqrt{M\omega/\pi\hbar}e^{-M \omega \boldsymbol{r}_{\textrm{CM}}^2/2\hbar }$, where $M=m_1+m_2+m_3$ is the total mass of the system.
However, the relative part of the wave function doesn't possess a simple expression as for the center-of-mass, since the corresponding Hamiltonian, i.e. $\mathcal{H}_{\rm{rel}}$, contains all the relevant potential terms.
In order to solve the Schr\"odinger equation of $\mathcal{H}_{\rm{rel}}$ we express the corresponding relative Jacobi vectors, i.e. ${\boldsymbol{\rho}}_1^{(k)}$ and $\boldsymbol{\rho}_2^{(k)}$, in the hyperspherical coordinates that consist of the hyperradius $R=\sqrt{(\boldsymbol{\rho}_1^{(k)})^2+(\boldsymbol{\rho}_2^{(k)})^2}$ and a set of hyperangles $\Omega^{(k)}$~\cite{rittenhouse_greens_2010,greene_clusters_2017}. 
The hyperradius $R$ indicates the entire system size whereas $\Omega^{(k)}$ collectively denotes the hyperangles which track the orientation of the particles on the 2D plane [Fig.~\ref{Fig:Jacobi}]. More specifically, $\Omega^{(k)}=\{\alpha^{(k)},\theta_1^{(k)},\theta_2^{(k)}\}$, where $\theta_j^{(k)}$ are the polar angles associated to the $\boldsymbol{\rho_j^{(k)}}$ Jacobi vectors, and $\alpha^{(k)}$ characterizes the length ratio between the two Jacobi vectors, i.e. $\rho_1^{(k)}=R\sin \alpha^{(k)}$ and $\rho_2^{(k)}=R \cos \alpha^{(k)}$. The resulting relative Hamiltonian in this coordinate system takes the following form:
\begin{eqnarray}
    &~&\mathcal{H}_{\rm{rel}}=-\frac{\hbar^2}{2\mu R^{3/2}}\frac{\partial^2}{\partial R^2} R^{3/2}+\frac{3\hbar^2}{8\mu R^2} + \frac{1}{2}\mu \omega^2 R^2+\mathcal{H}_{\rm{ad}}(R;\Omega), \nonumber \\ &~&\label{Eq:Hamilt_hyper} \\
    &~& \mathcal{H}_{\rm{ad}}(R;\Omega)=\frac{\hbar^2 \boldsymbol{\hat{\Lambda}}^2}{2\mu R^2}+\sum_{i<j} V_{ij}(R;\Omega^{(k)}),
    \label{Eq:Hamilt_adiabat}
\end{eqnarray}
where in Eq. (\ref{Eq:Hamilt_hyper}) the first three terms depend only on the hyperradius $R$, denoting the kinetic term and the trapping potential, respectively.
In Eq. (\ref{Eq:Hamilt_adiabat}) the first term of $\mathcal{H}_{\rm{ad}}(R;\Omega)$ describes the centrifugal motion of the three particles where the hyperangular operator, $\boldsymbol{\hat{\Lambda}}^2$, contains all the hyperangles $\Omega^{(k)}$, expressed in any of the three possible configurations $k=1,2,3$ ~\cite{smirnov_method_1977,avery_hyperspherical_1989}. 
The second term of Eq. (\ref{Eq:Hamilt_adiabat}) refers to the three pairwise interactions which couple the hyperradial and hyperangular degrees of freedom.

In order to solve the corresponding three-body Schr\"odinger equation of Eq. (\ref{Eq:Hamilt_hyper}) we choose the relative three-body wave function to obey the ansatz $\Psi(R,\Omega)=R^{-3/2} \sum_{\nu} F_{\nu}(R) \Phi_{\nu}(R;\Omega)$, where the hyperradius $R$ is treated as an adiabatic parameter.
This is the so-called adiabatic hyperspherical representation, where $F_\nu(R)$ and $\Phi_{\nu}(R;\Omega)$ denote the $\nu-$th hyperradial and hyperangular part of $\Psi(R,\Omega)$, respectively. 
In particular, the hyperangular $\Phi_{\nu}(R;\Omega)$ is obtained by diagonalizing Eq. (\ref{Eq:Hamilt_adiabat}) at fixed R. 
Namely, the corresponding fixed-R hyperangular Schr\"odinger equation reads:
\begin{equation}
  \left[\frac{2 \mu R^2}{\hbar^2}\mathcal{H}_{\rm{ad}}(R;\Omega)-(s^2_{\nu}(R)-1) \right]\Phi_{\nu}(R;\Omega)=0,
     \label{Eq:Hyperangular}
\end{equation}
where $s_{\nu}(R)$ indicate the eigenvalues of $\mathcal{H}_{\rm{ad}}(R;\Omega)$ for fixed $R$.
Note that, in the following, for notation simplicity we drop the $R$ dependence from the $s_{\nu}(R)\equiv s_{\nu}$ eigenvalues.
In order to tackle the hyperangular Schr\"odinger equation we exploit the fact that the two-body interactions are $\delta-$functions pseudopotentials. 
This allows us to semi-analytically solve \cref{Eq:Hyperangular} by employing the Green's function method \cite{rittenhouse_greens_2010} together with the corresponding two-body boundary conditions [for details see also Appendix \ref{Ap:Hyperangles}]. 
Under these considerations, the hyperangular eigenfunction $\Phi_{\nu}(R;\Omega^{(k')})$ for the $k'-$th Jacobi tree of Eq. \eqref{Eq:Hyperangular} takes the form:

\begin{eqnarray}
  \Phi_{\nu}(R;\Omega^{(k')})&=& - \sum_{k=1}^{3}\sum_{l=\pm L} C_{\nu,l}^{(k)}(R)Y_l(\theta_2^{(k)}) Y_0(\theta_1^{(k)}) \cos^{\abs{l}} \alpha^{(k)}\nonumber \\ & & \times _2F_1\left(\frac{\tilde{l}-s_{\nu}}{2},\frac{s_{\nu}+\tilde{l}}{2};\tilde{l};\cos^2(\alpha^{(k)}) \right) \nonumber \\
  & & \times \frac{\Gamma\left( \frac{s_{\nu}+\tilde{l}}{2}\right)\Gamma\left(\frac{\tilde{l}-s_{\nu}}{2}\right)}{2\Gamma(\tilde{l})}, \label{Eq:Hyperangular_analytic}
\end{eqnarray}
where $_2F_1(a,b;c;z)$ is the Gauss hypergeometric function, $\Gamma(\cdot)$ is the gamma function \cite{abramowitz_handbook_1965}, $\tilde{l}=\abs{l}+1$, $Y_l(\theta_2^{(k)})=\frac{e^{i l \theta_2^{(k)}}}{\sqrt{2\pi}}$ are plane-waves and $L$ denotes the total angular momentum carried by the system. The sum is over the three possible Jacobi trees $k=1,2,3$ [see also Fig. \ref{Fig:Jacobi}], each weighted by the coefficients $C_{\nu,l}^{(k)}(R)$. The three Jacobi trees are connected to the specific $k'$ tree on the left hand side of Eq. \eqref{Eq:Hyperangular_analytic} via a set of geometric relations \cite{nielsen_three-body_2001}.
It should be noted that by interrelating the $C_{\nu,l}^{(k)}(R)$ coefficients of different $k$ the bosonic or fermionic character of the particles can be specified [see also discussion below].

By utilizing the analytic expression for $\Phi_{\nu}(R;\Omega^{(k)})$ and the two-body boundary conditions we obtain a matrix equation for the $C_{\nu,l}^{(k)}(R)$ coefficients [for details see also Appendix \ref{Ap:Hyperangles}], which reads
\begin{eqnarray}
& & \sum_k M^l_{k'k}C_{\nu,l}^{(k)}(R)=0,  \label{Eq:Trans}  \\
   & & M^l_{k'k}= \begin{cases}   \ln\left(\frac{d_{k}Re^{-\gamma}}{a^{(k)}}\right)-\frac{1}{2}\psi\left(\frac{\tilde{l}-s_{\nu}}{2}\right)-\frac{1}{2}\psi\left(\frac{\tilde{l}+s_{\nu}}{2} \right), \: k'=k \\
    (-1)^l\frac{\Gamma\left(\frac{s_{\nu}+\tilde{l}}{2}\right)\Gamma\left(\frac{\tilde{l}-s_{\nu}}{2}\right)}{2\Gamma(\tilde{l})}f(\beta_{k'k}), \: k'\neq k, \label{Eq:matrix_elem}
   \end{cases}\\
  & & f(\beta_{k'k})=\cos^{\abs{l}}(\beta_{k'k})_2F_1\left(\frac{\tilde{l}-s_{\nu}}{2},\frac{s_{\nu}+\tilde{l}}{2};\tilde{l};\cos^2\beta_{k'k} \right),
\end{eqnarray}
where $\beta_{k'k}=\arctan \left[ \frac{(m_1+m_2+m_3)\mu}{m_km_{k'}} \right]$, $\tilde{l}=\abs{l}+1$, and $\psi(\cdot)$ is the digamma function.
The hyperangular eigenvalues $s_{\nu}$ are obtained by searching for zero-eigenvalues of the matrix $\boldsymbol{M^l}$ at fixed $R$ whereas the elements of the corresponding eigenvector determine the $C_{\nu,l}^{(k)}(R)$ coefficients \cite{rittenhouse_greens_2010,Incao_recomb2D_2015,nielsen_three-body_2001}.

\begin{table}
\setlength\extrarowheight{4pt}
\setlength{\tabcolsep}{5pt}
    \centering
    \begin{tabular}{|c|c|c|c|c|c|c|}
    \hline   $\textrm{ABC}$ &  $C_{\nu,l}^{(1)}$ & $a^{(1)}$ & $C_{\nu,l}^{(2)}$ & $a^{(2)}$ & $C_{\nu,l}^{(3)}$ & $a^{(3)}$    \\ \hline
    $\textrm{FFX}$   &  $C_{\nu,\pm 1}^{FX}$ & $a_{FX}$  & $-C_{\nu,\pm 1}^{FX}$ & $a_{FX}$  & $0$ & $0$ \\ \hline
    $\textrm{BBX}$ & $C_{\nu,0}^{BX}$ & $a_{BX}$ & $C_{\nu,0}^{BX}$ & $a_{BX}$ &  $C_{\nu,0}^{BB}$ & $a_{BB}$ \\ \hline
    \end{tabular}
    \caption{The interrelation of $C_{\nu,l}^{(k)}$ coefficients due to particle symmetry and the corresponding scattering lengths $a^{(k)}$ in the case of two identical spin-polarized fermions (FFX) and two bosons (BBX), together with a third distinguishable particle. The table also connects the odd-man-out and the descriptive notation.}
    \label{Tab:Symmetries}
\end{table}

In Eqs. (\ref{Eq:Trans}) and (\ref{Eq:matrix_elem}) the particle symmetry is not specified and in principle refer to three-body systems where all particles are distinguishable with each other.
As we mentioned at the beginning of this section we are primarily interested in three-body systems where two particles are identical obeying either bosonic or fermionic symmetry and the third one is distinguishable.
Eqs. (\ref{Eq:Trans}) and (\ref{Eq:matrix_elem}) in order to be symmetry adapted, additional constraints on the $C_{\nu,l}^{(k)}$ coefficients must be imposed.
Table \ref{Tab:Symmetries} shows the symmetry adapted $C_{\nu,l}^{(k)}(R)$ coefficients and the corresponding notation for the scattering lengths $a^{(k)}$ of two spin-polarized fermions (FF) or two identical bosons (BB) interacting with a third distinguishable atom (X). 
For example, if the particles $(1)$ and $(2)$ are identical fermions, then only four coefficients are non-zero, namely $C_{\nu,\pm 1}^{(1)}=-C_{\nu, \pm 1}^{(2)}=C_{\nu, \pm 1}^{FX}$, and the third coefficient being zero due to the lack of $p$-wave interactions. 
The latter also implies that the $s-$wave interaction between the two fermions is zero, thus $a_{FF}=0$, yielding thus one scattering length, $a_{FX}$, which describes the interaction between each fermion with the distinguishable atom [see also Table \ref{Tab:Symmetries}].

Using the eigenvalues $s_{\nu}$ and the eigenfunctions $\Phi_{\nu}(R;\Omega)$ of \cref{Eq:Hyperangular}, in the three-body Schr\"odinger equation belonging to the Hamiltonian $\mathcal{H}_{\rm{rel}}$, and by integrating over all the hyperangular degrees of freedom, a system of coupled one-dimensional ordinary differential equations for the hyperradial degree-of-freedom is obtained.
\begin{eqnarray}
  & &  \Big\{ -\frac{\hbar^2}{2\mu}\frac{d^2}{d R^2}+U_{\nu}(R)  \Big\} F_{\nu}(R) \nonumber \\
& &    -\frac{\hbar^2}{2\mu} \sum_{\nu'} \left[2P_{\nu \nu'}(R) \frac{d}{dR}+Q_{\nu \nu'}(R) \right] \,F_{\nu'}(R)=EF_{\nu}(R). \nonumber \\
\label{Eq:Hyperradial}
\end{eqnarray}
Here, $F_{\nu}(R)$ is the hyperradial part of the relative three-body wave function, $U_{\nu}(R)$ indicates the $\nu$-th adiabatic potential that includes the trap, whereas the terms $P_{\nu \nu'}(R)$ and $Q_{\nu \nu'}(R)$ denote the non-adiabatic coupling matrix elements \cite{greene_clusters_2017}.
More specifically, $U_{\nu}(R)$, $P_{\nu \nu'}(R)$ and $Q_{\nu \nu'}(R)$ are given by the following expressions:
\begin{eqnarray}
  U_{\nu}(R)&=&\frac{\hbar^2}{2\mu R^2}\left(s_{\nu}^2-\frac{1}{4} \right)+\frac{1}{2}\mu \omega^2 R^2, \label{Eq:Pots_adiabatic}  \\
  P_{\nu \nu'}(R)&=&\braket{\Phi_{\nu}(R;\Omega^{(k)})|\frac{\partial \Phi_{\nu'}(R;\Omega^{(k)})}{\partial R}}_{\Omega},  \label{Eq:P_coup}\\
  Q_{\nu \nu'}(R)&=&\braket{\Phi_{\nu}(R;\Omega^{(k)})|\frac{\partial^2 \Phi_{\nu'}(R;\Omega^{(k)})}{\partial R^2}}_{\Omega}, \label{Eq:Q_coup}
\end{eqnarray}
where the symbol $\braket{\ldots}_{\Omega}$ indicates that the integration is over the hyperangles only.
Due to the zero-range interactions the $P_{\nu \nu'}(R)$ and $Q_{\nu \nu'}(R)$ matrix elements have semi-analytical expressions as shown in Refs. ~\cite{rittenhouse_greens_2010,kartavtsev_universal_2006,kartavtsev_universal_2007}.
This particular feature simplifies the numerical diagonalization of Eq. (\ref{Eq:Hyperradial}) where the hyperradial solutions $F_{\nu}(R)$, which are expanded in the basis of B-splines \cite{deBoor_practical_1978}, obey the vanishing boundary conditions at the origin and asymptotically. 
The former is a result of the repulsive nature of the $U_\nu(R)$ potentials at short hyperradii and the latter occurs due to the 2D harmonic trap.
We should note that in the following sections and in the appendices the notation $F^j_{\nu}(R)$ and $E^j$ signifies the $j-$th eigenvector and eigenvalue of Eq. (\ref{Eq:Hyperradial}) respectively.

\begin{figure*}[t]
    \centering
    \includegraphics[width=1 \textwidth]{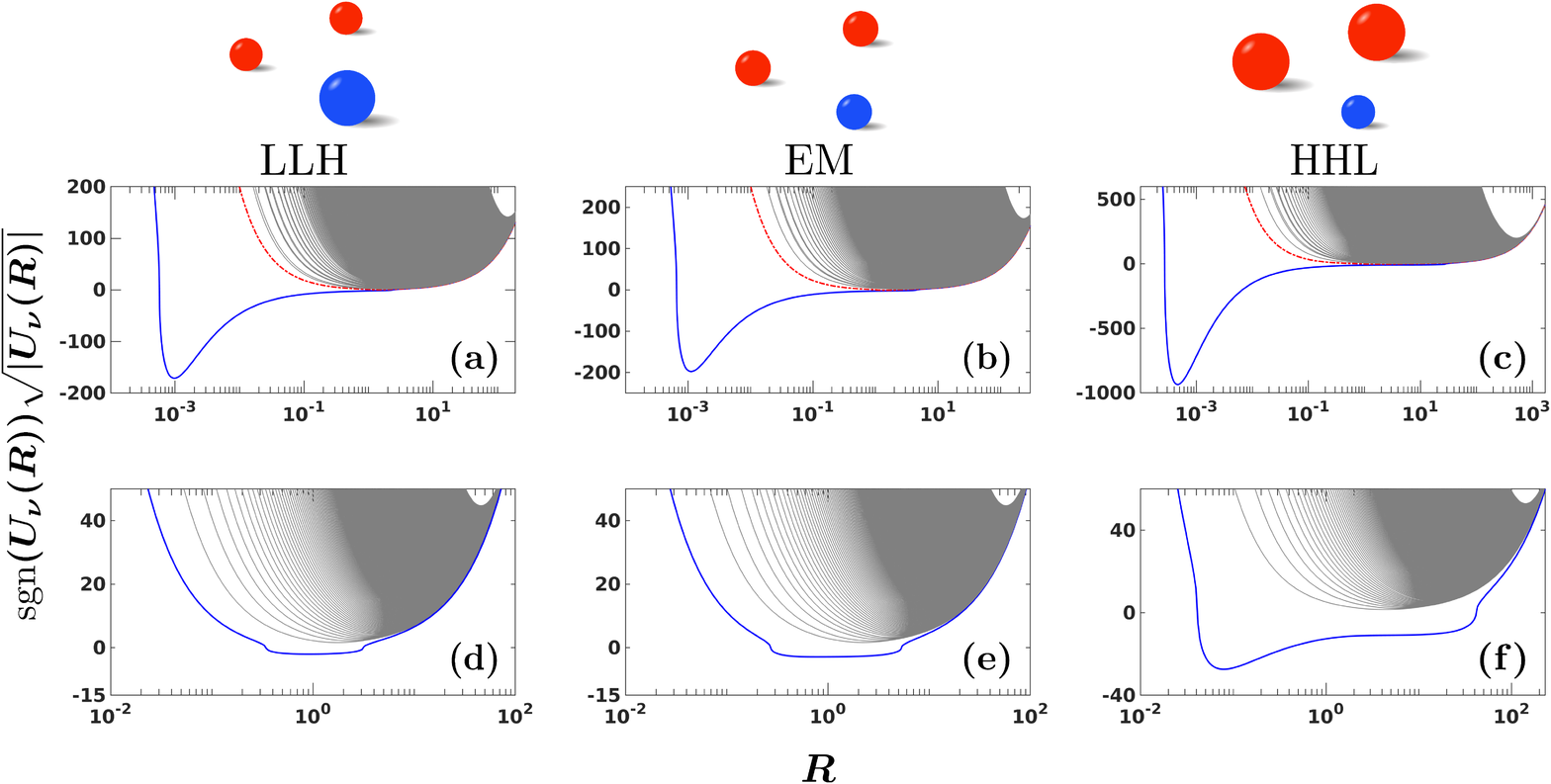}
    \caption{Rescaled adiabatic potentials $\textrm{sgn}(U_{\nu}(R))\sqrt{|U_{\nu}(R)|}$ of (a)-(c) BBX and (d)-(f) FFX three-body mixtures for different hyperadius $R$. The LLH to HHL transition occurs from left to right as $\mathcal{M}=\frac{m_{B/F}}{m_X}$ increases taking values $m_B/m_X=0.04, 1.16, 22.16$, and $m_F/m_X=0.045, 0.86, 24.71$.
    The identical particles (denoted with red circles) become heavier than the distinguishable particle (marked by blue circles). The corresponding 2D scattering lengths obey (a)-(c) $1/a_{FX}=2.7$, and (d)-(f) $a_{BB}/a_{BX}=2$. Harmonic oscillator units are employed.}
    \label{Fig:Pots}
\end{figure*}

In the following, we will mainly focus on two different types of three-body mixture systems. 
The first comprises two spin-polarized fermions interacting with a distinguishable particle. 
It will be termed FFX and exhibits total angular momentum $L=1$ with an antisymmetric wave function upon exchange of the two identical fermions i.e., $L^{\pi}=1^-$, where $\pi$ is the total parity of the system, a mass ratio $\mathcal{M}=m_F/m_X$ and scattering length $a_{FX}$. 
The second setup consists of two identical interacting bosons coupled with a third atom. 
This setting is dubbed BBX and it is characterized by  $L^{\pi}=0^+$, a mass ratio $\mathcal{M}=m_B/m_X$ as well as two scattering lengths $a_{BB}$ and $a_{BX}$ for the identical bosons and between the two different species respectively. For convenience, in the following, we will consider their ratio $a_{BB}/a_{BX}$ as the relevant interaction parameter. Owing to the spherical symmetry of the interactions the total angular momentum $L$ is conserved. However, the two-body angular momenta $(l_1,l_2)$ that construct the $L-$space are also decoupled in our case because we have only considered $s$-wave interactions \cite{Volosniev_Borromean_2014}.
Hereafter, harmonic oscillator units are adopted, meaning that $\hbar=\omega=m_{B/F}=1$, where $m_{B/F}$ denotes the mass of the majority species (two identical atoms), unless it is specified otherwise. 
It is also worth noting that considering a radial trapping frequency of $\omega= 2\pi \, \times \, 20 \,\textrm{Hz}$, typical in 2D experiments \cite{jochim_anomalous_2018,jochim_anomalous_2019}, then the harmonic oscillator length takes the values $a_0=4.58, 9.22$ $\mu \textrm{m}$ when $^{173}\textrm{Yb}-^{173}\textrm{Yb}-^7\textrm{Li}$ and $^6\textrm{Li}-^6\textrm{Li}-^{133}\textrm{Cs}$ FFX settings are considered. Similarly, $a_0$ takes the values $a_0=5.29, 8.53 \, \mu \textrm{m}$ for $^{133}\textrm{Cs}-^{133}\textrm{Cs}-^6\textrm{Li}$ and $^7\textrm{Li}-^7\textrm{Li}-^{173}\textrm{Yb}$ BBX systems respectively.

\section{Adiabatic hyperspherical potentials and energy spectra} \label{Sec:Spectra}

\begin{figure*}[t]
    \centering
    \includegraphics[width=1 \textwidth]{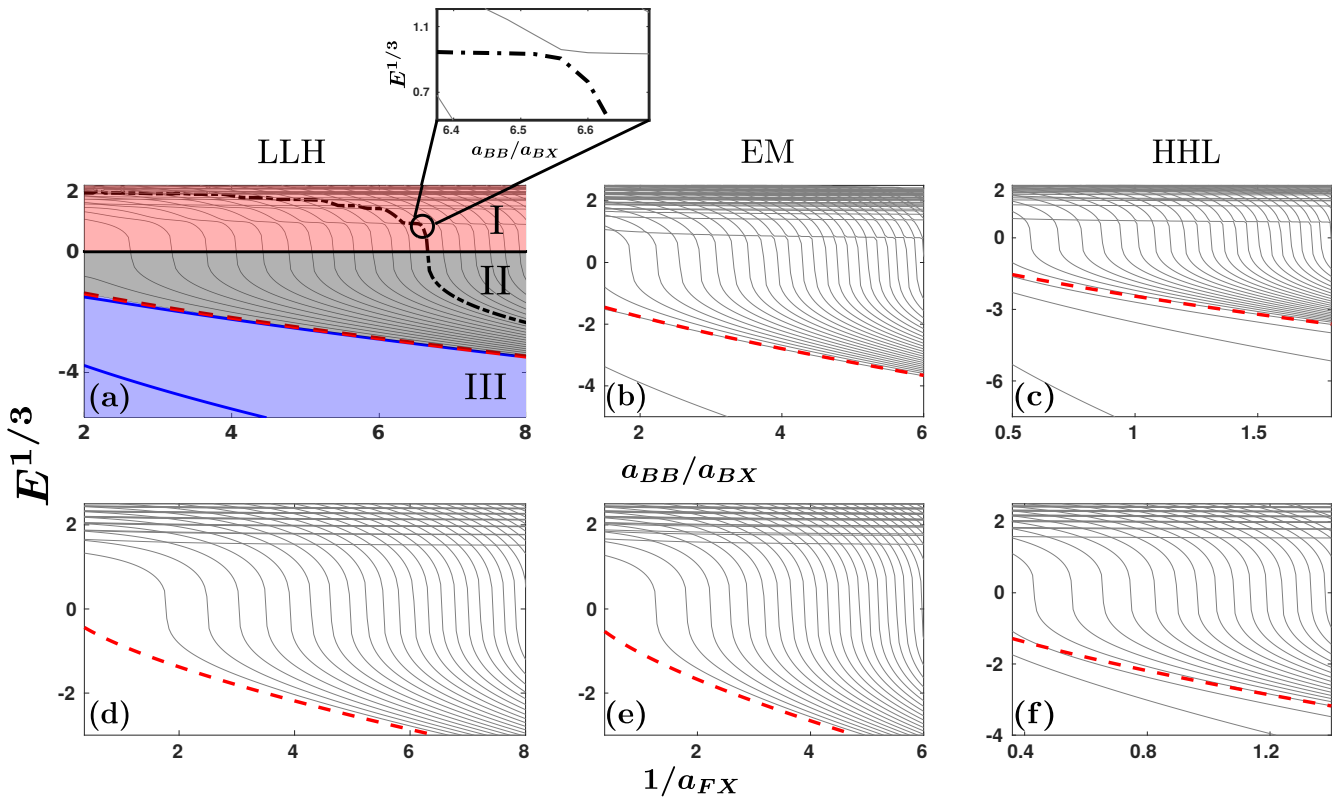}
    \caption{Rescaled energy spectra $E^{1/3}$ of (a)-(c) BBX and (d)-(f) FFX systems for typical LLH, EM and HHL cases (see legends) characterized by mass ratio $m_B/m_X=0.04, 1.16, 22.16$ and $m_F/m_X=0.045, 0.86, 24.71$ respectively for varying scattering length. In all panels, the red dashed line indicates the atom-dimer threshold (a)-(c) $E_{BX}$ and (d)-(f) $E_{FX}$ separating the trimer from the atom-dimer and trap states. 
    A schematic illustration of the aforementioned energy regions in which these states occur is provided in panel (a) where the black solid horizontal line separates the trap from the atom-dimer states. Particularly, type-I mark the trap and atom-dimer states, type-II the atom-dimer and type-III indicate the trimer states. The type-I trap and atom-dimer states change character in the vicinity of avoided-crossings (see for instance the circle). The inset in panel (a) showcases in detail such an avoided crossing.  
    All quantities are expressed in harmonic oscillator units.}
    \label{Fig:Spectra}
\end{figure*}

\begin{table}[t]
\setlength\extrarowheight{5pt}
    \centering
    \begin{tabular}{|c|c|c|c|}
    \hline $\textrm{ABC}$  & LLH   &  EM & HHL  \\  \hline
     $\textrm{BBX}$ & $^7\textrm{Li}-^7\textrm{Li}-^{173}\textrm{Yb}$ &  $^7\textrm{Li}-^7\textrm{Li}-^6\textrm{Li}$ &  $^{133}\textrm{Cs}-^{133}\textrm{Cs}-^6\textrm{Li}$ \\ \hline
     $\textrm{FFX}$ & $^6\textrm{Li}-^6\textrm{Li}-^{133}\textrm{Cs}$ & $^6\textrm{Li}-^6\textrm{Li}-^7\textrm{Li}$ & $^{173}\textrm{Yb}-^{173}\textrm{Yb}-^7\textrm{Li}$ \\ \hline
    \end{tabular}
    \caption{Representative cases of identical particles with fermionic and bosonic symmetry, which are lighter (LLH), roughly equal in mass (EM) and heavier than a third distinguishable particle (X).}
    \label{Tab:Systems}
\end{table}

To shed light into the eigenspectrum of three-body mixtures and their microscopically allowed processes, in the following, we investigate the potential curves and the corresponding hyperradial spectrum for the FFX and BBX systems. 
More specifically, we consider three representative cases of different mass ratio, where the two identical particles, with either bosonic or fermionic symmetry, are lighter, roughly equal in mass, and heavier than the third distinguishable particle.
These three distinct scenarios are referred to as light-light-heavy (LLH), equal-massed (EM) and heavy-heavy-light (HHL), respectively~\cite{greene_clusters_2017}.
The adiabatic potential curves obtained via Eq. \eqref{Eq:Pots_adiabatic} for the above-described settings are depicted in Fig.~\ref{Fig:Pots}. 
Note that the selection of the specific mass ratio corresponds to the experimentally relevant atomic species reported in Table \ref{Tab:Systems}. 

The two energetically lowest potential curves, $U_1(R)$ (blue solid line) and $U_2(R)$ (red dash-dotted line), shown in \cref{Fig:Pots} (a)-(c) represent the ones which in the absence of a trap approach asymptotically, i.e. $R \to \infty$, the BX+B and BB+X atom-dimer thresholds. The latter have energy $E_{BX}=-2e^{-2\gamma}(1+\mathcal{M})/a_{BX}^2$ and $E_{BB}=-4e^{-2\gamma}/a_{BB}^2$ respectively \cite{Incao_recomb2D_2015}. 
However, here at large hyperradii the harmonic 2D trap dominates and thus these two potential curves coincide scaling as $\sim R^2$  \cite{rittenhouse_hyperspherical_2016,daily_resonances_2010}.
In the limit of small hyperradius $R$ (i.e. $R \to 0$),  $U_1(R)$ and $U_2(R)$ exhibit a repulsive potential "wall" preventing in this manner the three atoms to approach together at short distances. 
Another notable feature of $U_1(R)$ (blue line) is that independently of the mass ratio it possesses a classically allowed region at small $R$ where the potential is deep enough in order to support trimer states.
The remaining gray solid lines in \cref{Fig:Pots}(a)-(c) represent potential curves, with high hyperangular momentum $s_\nu$, which describe the effective centrifugal forces between the three atoms \cite{Incao_recomb2D_2015}.

For FFX systems, the corresponding potential curves [see \cref{Fig:Pots} (d)-(f)] exhibit significantly altered characteristics from the BBX setting. Indeed, there is only one atom-dimer threshold, i.e. FX+F, since the two identical fermions do not interact. 
The associated potential curve is illustrated in panels (d)-(f) by the blue solid line.
We remark that in the absence of a trapping potential this potential curve asymptotically saturates at an energy $E_{FX}=-2e^{-2\gamma}(1+\mathcal{M})/a_{FX}^2$ \cite{Incao_recomb2D_2015}. 
Moreover, in the classically allowed region, the lowest potential is not deep enough to maintain trimer states for all mass ratios. 
Namely, only for systems with an adequately large mass ratio (HHL), the lower potential curve possesses a pronounced well [\cref{Fig:Pots} (f)] in contrast to BBX systems, where a deep well is always present [\cref{Fig:Pots} (a)-(c)]. 

The hyperradial spectra of the potential curves [\cref{Fig:Pots}] are provided in \cref{Fig:Spectra}.
Namely, panels in \cref{Fig:Spectra} (a)-(c) [(d)-(f)] refer to the adiabatic potential curves depicted in \cref{Fig:Pots} (a)-(c) [(d)-(f)] and corresponding to the BBX [FFX] system at three different mass ratios.
Evidently, 
three types of bound states are discernible: I) trap states and atom-dimer states with energies $E >0$, II) purely atom-dimer states (black dash-dotted line)  with dimer energies $E_{\sigma \sigma'}\le E<0$, where $\sigma=~B,~F,~X$ and $\sigma \neq \sigma'$, and III) trimer bound states (blue solid line) with energies $E<E_{\sigma \sigma'}$. 
Note that the red dashed lines in \cref{Fig:Spectra} denote the $\sigma \sigma'$-dimer energies $E_{\sigma \sigma'}$, whereas the black solid horizontal line in panel (a) depicts $E=0$.
The trap states of type-I [Fig. \ref{Fig:Spectra} (a)] correspond to three weakly interacting trapped atoms. 
They emerge at $E>0$ and are virtually independent of $a_{BB}/a_{BX}$ or $1/a_{FX}$ as depicted in Fig. \ref{Fig:Spectra} (a) ~\cite{rittenhouse_hyperspherical_2016}. 
It is also important to mention that the energetically lowest trap state in BBX sytems takes place at energies $E>0$ whereas for FFX they emerge for $E>1$.
The type-II states are associated with the formation of an atom and a dimer, while their energy lies between the dimer energy and 0.
In the limit of $a_{BB}/a_{BX} \gg 1$ [$1/a_{FX} \gg 1$], the atom-dimer energies behave like $-a_{BB}/a_{BX}^2$ ($-1/a_{FX}^2$) for BBX [FFX], see in particular Fig. \ref{Fig:Spectra} (a) [(d)] \cite{rittenhouse_hyperspherical_2016,liu_correlated2D_2010}.
For large $1/a_{FX}$ and $a_{BB}/a_{BX}$, the energy difference of two successive eigenstates, approaches $\Delta E=2$ \cite{portegies_trap_2011}, which is affected due to the employed scaling in Fig. \ref{Fig:Spectra}. It is the excitation energy of the particle accompanying the dimer and stems from the harmonic trap.  
Due to the coupling between different adiabatic potentials via the $P$ and $Q$ non-adiabatic elements [Eqs. \eqref{Eq:P_coup}, \eqref{Eq:Q_coup}], the trap states change character in the vicinity of avoided-crossings, alternating between type-I atom-dimer and energetically lower trap states [see the circle in Fig. \ref{Fig:Spectra} (a)]. 
The distinction between these two states in region I is more prominent in the case of sharp avoided-crossings [e.g. at $a_{BB}/a_{BX} \simeq 7$ in Fig. \ref{Fig:Spectra} (a)], compared to the case of broad avoided-crossings [e.g. at $a_{BB}/a_{BX} \simeq 3$ in Fig. \ref{Fig:Spectra} (a)].

\begin{figure}[t]
    \centering
    \includegraphics[width=0.5 \textwidth]{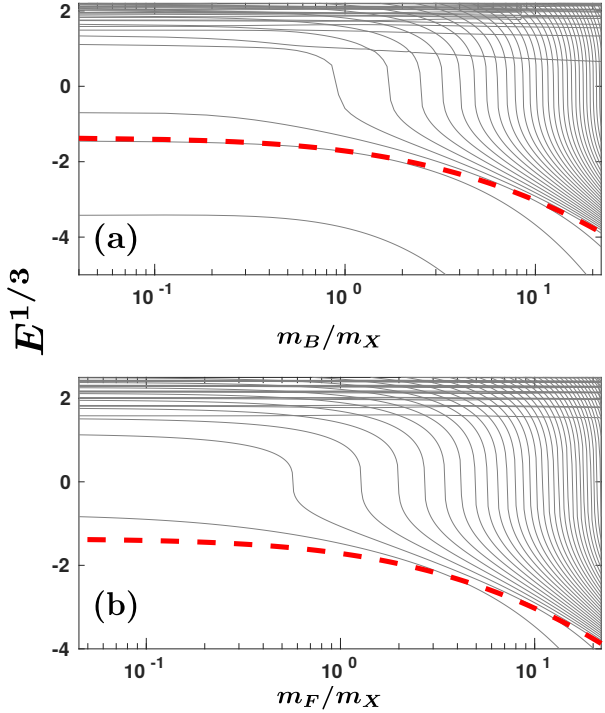}
    \caption{Rescaled energy spectrum $E^{1/3}$ for (a) BBX and (b) FFX systems with respect to the mass ratio $m_B/m_X$ and $m_F/m_X$ respectively. The considered scattering length ratios are (a) $a_{BB}/a_{BX}=2$ and (b) $1/a_{FX}=2$. The red dashed line denotes the energy of the dimer in each case, namely $E_{\sigma X}=-8e^{-2\gamma}(1+\mathcal{M})$, where $\sigma=B/F$ and $\mathcal{M}=\frac{m_{B/F}}{m_X}$.}
    \label{Fig:Spectra_mass}
\end{figure}

The type-III states are related to trimers which energetically occur below the dimer energy $E_{\sigma \sigma'}$.
More specifically, for the BBX system we observe in \cref{Fig:Spectra} (a)-(c) that as the mass ratio increases the number of trimer states ranges from 2 to 4. 
This is an immediate effect of the corresponding potential curve, i.e. $U_1(R)$ in \cref{Fig:Pots} (a)-(c) which deepens as we transition from a LLH scenario to a HHL one \cite{sandoval_radii2D_2016}.
On the other hand, regarding the FFX system trimer states are visible only for a HHL case [\cref{Fig:Spectra} (f)].
The aforementioned aspects of the 2D three-body collisions can be clearly inferred by inspecting \cref{Fig:Spectra_mass}. 
As can be seen, for the BBX system [\cref{Fig:Spectra_mass} (a)] at fixed $a_{BB}/a_{BX}=2$ there is at least one trimer state at {\it any} mass ratio, whilst for the FFX system [\cref{Fig:Spectra_mass} (b)] there are no trimer states at least within the regime of light identical fermions and a heavy spectator particle \cite{pricoupenko_planar_2010}.
 Indeed, a detailed calculation of the eigenvalues of Eq. \eqref{Eq:Hyperradial} explicates that for a scattering length $1/a_{FX}=2$ the first trimer state occurs at a critical mass ratio $\mathcal{M}^*=3.817$.
 This value is larger than $\mathcal{M}^*=3.34$ reported in Ref. ~\cite{pricoupenko_planar_2010}, which treats the FFX system in the absence of a trap. 
 Note that a similar effect occurs in 3D FFX systems where the first trimer state appears at even larger mass ratio, i.e. $\mathcal{M}^*=8.17$ \cite{kartavtsev_low-energy_2007}.
 The fact that trimer states emerge at larger mass ratio compared to free space can be explained via the behavior of the first adiabatic potential, $U_1(R)$ [Fig. \ref{Fig:Pots} (d)]. The trapping potential contribution [second term in Eq. \eqref{Eq:Pots_adiabatic}] shifts $U_1(R)$ to more positive values, compared to the adiabatic potential term [first term in Eq.~\eqref{Eq:Pots_adiabatic}] in free space. Accordingly, $U_1(R)$ becomes shallower in the presence of a trap and therefore larger mass ratios deepen $U_1(R)$, favoring in turn the formation of trimer states.
 
\section{Few-body correlations and Tan contacts} \label{Sec:Contacts}

Few-body correlations of short-range interacting atomic ensembles are embedded in the momentum distribution of the reduced one-body density in the limit of large momenta~\cite{tan_contact1_2008,tan_contact2_2008,tan_contact3_2008,castin_general_2012}. 
This observable can be routinely measured in ultracold atom experiments via time-of-flight measurements ~\cite{stewart_contact_2010,sagi_measurement_2012}.
For binary three-body mixtures this asymptotic expansion is related to the relevant short-range two- and three-body correlation functions~\cite{castin_general_2012} via the so-called two- and three-body contacts ~\cite{braaten_universal_2011,Bellotti_contacts_2014}. 
In particular, the two-body contact has been directly measured via radio-frequency spectroscopy \cite{wild_measurement_2012,sagi_measurement_2012}.
More specifically, in the limit of large momenta the momentum distribution of the  $\sigma=B/F$ or $X$ species reduced one-body density [see also \cref{Ap:Hyperangles} and \cref{Ap:Asymptotic}] takes the form

\begin{equation}
n_{\sigma}(\boldsymbol{p}_{\sigma}) \approx  n^a_{\sigma}(\boldsymbol{p}_{\sigma})+  n^b_{\sigma}(\boldsymbol{p}_{\sigma}), \label{Eq:Asymptotic_expansion_1}
\end{equation}
where the single-particle momentum of the $\sigma$ species $p_{\sigma}$ is larger than all the relevant momentum scales provided by the inverse scattering lengths $1/a_{\sigma\sigma'}$. Note that $\sigma=\sigma'$ ($\sigma'\neq \sigma$) denote the intraspecies (interspecies) interactions.  
\cref{Eq:Asymptotic_expansion_1} shows that the momentum distribution of the one-body density possesses two main contributions, namely $n^a_{\sigma}(\boldsymbol{p}_{\sigma})$ and $n^b_{\sigma}(\boldsymbol{p}_{\sigma})$ which are attributed to the presence of two-~\cite{castin_general_2012} and three-body correlations ~\cite{Bellotti_contacts_2014,bellotti_dimeffects_2013} respectively. 

\begin{figure*}[t]
    \centering
    \includegraphics[width=1 \textwidth]{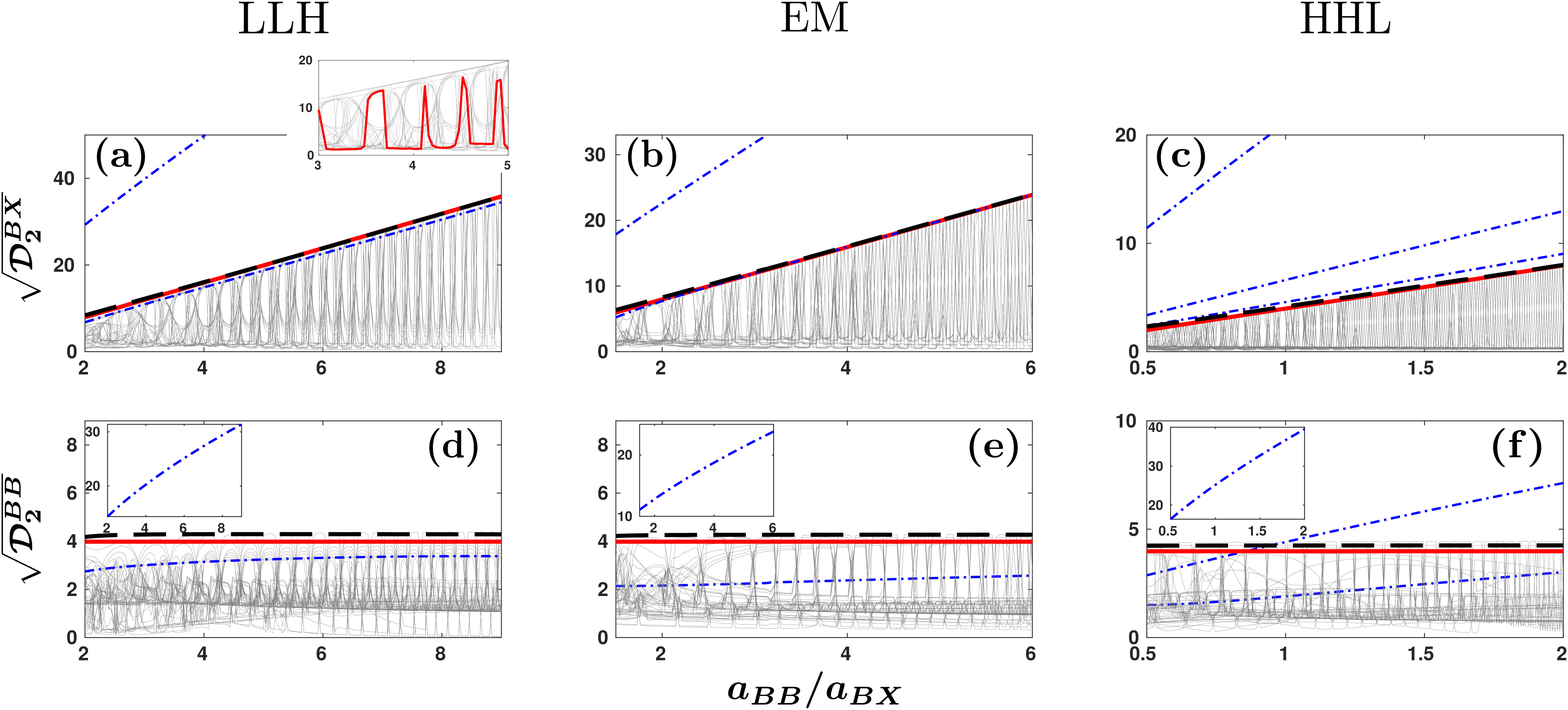}
    \caption{Two-body contact $\sqrt{\mathcal{D}_2^{\sigma\sigma'}}$ among the (a)-(c) BX and (d)-(f) BB species for three body settings ranging from LLH ($m_B/m_X=0.04$), EM ($m_B/m_X=1.16$) to HHL ($m_B/m_X=22.16$) as a function of $a_{BB}/a_{BX}$ [see also Table \ref{Tab:Systems}]. The inset in panel (a) displays exemplarily the oscillatory behavior of $\mathcal{D}_2^{BX}$ due to the change of character of type-I atom-dimer and trap states in the vicinity of avoided-crossings. The insets in panels (d)-(f) feature $\sqrt{\mathcal{D}_2^{BB}}$ of the first trimer state. The bound stemming from the analytical expression described by Eq. \eqref{Eq:Upper_bound}, is shown with the red solid line, whereas the upper bound stemming from the JWKB method is depicted by the black dashed line. The two-body contact of (non) trimer states is illustrated with the (gray solid) blue dash-dotted lines.}
    \label{Fig:Cont2_scattering_BBX}
\end{figure*}

More specifically, $n^a_{\sigma}(\boldsymbol{p}_{\sigma})$ contains terms solely associated with intra- and interspecies two-body correlations of the $\sigma$ species atom [see also Appendix \ref{Ap:Asymptotic}].
For instance, the asymptotic expansion of the reduced density of B species involves both intra- and interspecies two-body correlations, whereas the density of X species involves only interspecies ones.
Utilizing the hyperspherical approach, $n^a_{\sigma}(\boldsymbol{p}_{\sigma})$ can be expressed in terms of the hyperradial solutions $F_{\nu}(R)$ and $C_{\nu}^{\sigma\sigma'}$ coefficients of the hyperangular part of the three-body wave function  in the descriptive notation [see also \cref{Ap:Asymptotic} and \cref{Tab:Symmetries}]
\begin{eqnarray}
n^a_{\sigma}(\boldsymbol{p}_{\sigma})& =& \frac{4\pi}{ \mu N_{\sigma} p_{\sigma}^4}\sum_{\sigma'}\mu_{\sigma\sigma'} \int_0^{\infty} \frac{dR}{R^2} \abs{\sum_{\nu} F_{\nu}(R) \sum_{l= \pm L} C_{\nu,l}^{\sigma\sigma'}(R)}^2 \nonumber \\
& = & \frac{1}{N_{\sigma}p_{\sigma}^4} \sum_{\sigma'} (1+\delta_{\sigma\sigma'}) \mathcal{D}^{\sigma\sigma'}_2. \label{Eq:Cont2}
\end{eqnarray}
Here, $\mu_{\sigma\sigma'}$ is the two-body reduced mass between two atoms of the same species ($\sigma=\sigma'$) or two atoms belonging to different species ($\sigma \neq \sigma'$), while $N_{\sigma}$ is the $\sigma$ species particle number. 
Importantly, $\mathcal{D}^{\sigma\sigma'}_2$ signifies the intra- ($\sigma=\sigma'$) or interspecies ($\sigma \neq \sigma'$) two-body contact~\cite{castin_general_2012}. 
On the other hand, the term $n^{b}_{\sigma}(\boldsymbol{p}_{\sigma})$ is related to the product of inter- and intraspecies two-body correlations of the $\sigma$ species particle giving rise to the three-body ones. 
In the hyperspherical framework $n^{b}_{\sigma}(\boldsymbol{p}_{\sigma})$  reads:

\begin{eqnarray}
     n^{b}_{\sigma}(\boldsymbol{p}_{\sigma})&=& \frac{4\pi}{N_{\sigma}p_{\sigma}^4} \sum_{\sigma'}\frac{\mu_{\sigma\sigma'}}{\mu}  \int_0^{\infty} \frac{dR}{R^2} \,\Bigg\{ J_0\left[\frac{p_{\sigma} R \sqrt{\mu_{\sigma\sigma'}}}{\sqrt{\mu}} \right](-1)^L   \nonumber  \\
    &~& +J_{2L}\left[\frac{p_{\sigma} R \sqrt{\mu_{\sigma\sigma'}}}{\sqrt{\mu}} \right](1-\delta_{0,L}) \Bigg \} \nonumber \\
    &~& \times \sum_{\sigma'' \neq \sigma'} \sum_{l= \pm L} \left(\sum_{\nu} F_{\nu}(R) C_{\nu,l}^{\sigma\sigma'}(R) \right) \nonumber \\
    &~& \times \left(\sum_{\nu'} F_{\nu'}(R) C_{\nu',l}^{\sigma' \sigma''}(R) \right)^*
    \label{Eq:Asymptotic_expansion_2}
\end{eqnarray}
where $J_{\nu}( \cdot)$ is the $\nu$-th Bessel function of the first kind and $L$ is the total angular momentum of the three-body system. 
Note that in contrast to the two-body term $n^a_{\sigma}(\boldsymbol{p}_{\sigma})$, the single-particle momentum $p_{\sigma}$ is also involved into the integration of $n^{b}_{\sigma}(\boldsymbol{p}_{\sigma})$. 
Hence, the scaling of the latter with the momentum is different than $1/p_{\sigma}^4$, see in particular the discussion in Sec.~\ref{3b_contact}. 

\subsection{Scaling behavior of two-body correlations}  \label{Sec:Cont2}

The presence of two-body short-range correlations in binary mixtures, is captured by $\mathcal{D}_2^{\sigma\sigma'}$ [Eq. \eqref{Eq:Cont2}], i.e. the $\sigma~\sigma'$ two-body contact.
Intuitively, the contact $\mathcal{D}_2^{\sigma\sigma'}$ can, in principle, exhibit an increasing tendency as the $\sigma~\sigma'$ pair of particles comes closer relatively to the third particle.
This enhancement of $\mathcal{D}_2^{\sigma\sigma'}$ signals that the three-body is dominated by strong two-body correlations.
Therefore, it is anticipated that $\mathcal{D}_2^{\sigma\sigma'}$ possesses distinctive characteristics with respect to the particular type of eigenstate of the three-body system, i.e. referring to trimer, type-I/II atom-dimer and trap states.

\begin{figure*}[t]
    \centering
    \includegraphics[width=1 \textwidth]{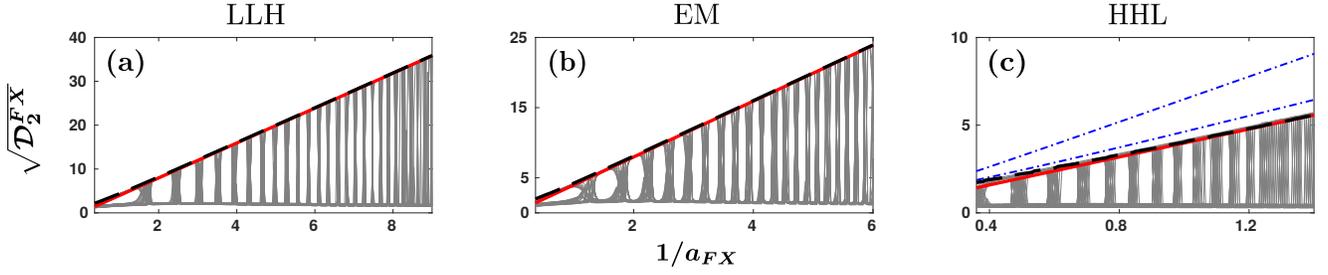}
    \caption{Two-body contact $\sqrt{\mathcal{D}_2^{FX}}$ between FX species in (a) LLH ($m_F/m_X=0.045$), (b) EM ($m_F/m_X=0.86$) and (c) HHL ($m_B/m_X=24.71$) cases for different $1/a_{FX}$. The analytical expression for the upper bound [Eq. \eqref{Eq:Upper_bound}] is shown with the red solid line, whereas the upper bound stemming from the JWKB method is denoted by the black dashed line. The two-body contact of (non) trimer states is showcased with (gray solid) blue dash-dotted lines.}
    \label{Fig:Cont2_scattering_FX}
\end{figure*}

For instance, the two-body contacts are strongly enhanced if the three particles are bounded in a trimer state, see in particular the blue dash-dotted lines in Fig. \ref{Fig:Cont2_scattering_BBX} and \ref{Fig:Cont2_scattering_FX} (c). 
Concretely, the two-body correlations of these states become substantial as $a_{BB}/a_{BX}$ [Fig. \ref{Fig:Cont2_scattering_BBX}] and $1/a_{FX}$ [Fig. \ref{Fig:Cont2_scattering_FX}] increase. 
This holds for both inter- ($\mathcal{D}_2^{BX}$) and intraspecies ($\mathcal{D}_2^{BB}$) correlations in BBX as well for $\mathcal{D}_2^{FX}$ in FFX settings. 
A general feature observed in both BBX [Fig. \ref{Fig:Cont2_scattering_BBX}] and FFX [Fig. \ref{Fig:Cont2_scattering_FX}] systems is that the two-body contact of all the other  eigenstates, that is atom-dimer and trap states showcased in gray lines, is confined within an envelope and oscillates between two values, see the red curve in the inset of Fig. \ref{Fig:Cont2_scattering_BBX} (a) and also the discussion below for more details.

The lower value of the two-body contact is associated with highly excited trap states. This lower bound depends on the energy of the aforementioned eigenstates and eventually tends to zero as energetically higher excited trap states are taken into account.
This is due to the fact that for highly excited trap states, the overall size of the three particles, as specified by the hyperradius $R$, increases compared to the size of the system residing in lower-lying energy states, yielding thus weak two-body correlations. This lower value of the contact is attained irrespectively  of the value of the scattering length $a_{BB}/a_{BX}$ and $1/a_{FX}$.
On the other hand, the oscillatory behavior of the $\mathcal{D}_2^{\sigma\sigma'}$, as demonstrated by the red curve in the inset of Fig. \ref{Fig:Cont2_scattering_BBX} (a), originates from the sharp avoided-crossings of the energy levels between the type-I atom-dimer and trap eigenstates [see the circle in Fig. \ref{Fig:Spectra} (a)].
Indeed, in the vicinity of these narrow avoided-crossings the spatial configuration of the three particles alters significantly, e.g. from a delocalized trap state into a type-I atom-dimer, where at most two particles are close to each other. 
Therefore, if the system configuration is that of an atom-dimer (trap state) it leads to an enhanced (reduced) contact due to the strong (weak) pair correlations. 
By tuning the scattering lengths $a_{BB}/a_{BX}$ and $1/a_{FX}$ towards the subsequent avoided-crossings this atom-dimer (trap) state becomes again a trap (atom-dimer) state, and $\mathcal{D}_2^{\sigma\sigma'}$ approaches once more its lower (upper) value. Notice that in Fig. \ref{Fig:Cont2_scattering_BBX} the two-body contact of the energetically lower fifty eigenstates is presented. A larger number of energy states results in the filling of the envelope by atom-dimer and trap states (gray lines) which exhibit an oscillatory behavior. Moreover, the lower bound of $\mathcal{D}_2^{\sigma\sigma'}$ has a value closer to zero compared to the case with fewer considered eigenstates. 

The upper value of the two-body contact is attributed to the presence of two-body correlations stemming from type-II purely atom-dimer states. 
In particular, the upper value of $\mathcal{D}_2^{BX}$ [Figs. \ref{Fig:Cont2_scattering_BBX} (a)-(c)] and $\mathcal{D}_2^{FX}$ [Figs. \ref{Fig:Cont2_scattering_FX}] becomes larger for increasing 
scattering lengths, $a_{BB}/a_{BX}$ and $1/a_{FX}$.
This is associated to the behavior of type-II purely atom-dimer states [see also the energy spectra in Fig. \ref{Fig:Spectra} (a)] whose energy increases in absolute value for larger scattering length. 
As a consequence, the dimer becomes strongly bound, leading to an enhanced $\mathcal{D}_2^{\sigma\sigma'}$. However, the upper value of $\mathcal{D}_2^{BB}$ 
[Figs. \ref{Fig:Cont2_scattering_BBX} (d)-(f)] remains almost constant when varying $a_{BB}/a_{BX}$. This occurs since $a_{BB}=1$.
The latter implies that the second adiabatic potential $U_2(R)$ [red dash-dotted line in Fig. \ref{Fig:Pots} (a)-(c)], which is associated with the BB+X atom-dimer threshold, is insensitive to variations of $a_{BB}/a_{BX}$. Hence the upper bound of $\mathcal{D}_2^{BB}$, being determined by 
the BB dimer states, is constant with respect to $a_{BB}/a_{BX}$. 
However, two-body correlations between the identical bosons are substantially enhanced for increasing $a_{BB}/a_{BX}$ when the BBX system resides in trimer states [blue dash-dotted lines in Figs. \ref{Fig:Cont2_scattering_BBX} (d)-(f)]. This becomes more prominent in the HHL case [Fig. \ref{Fig:Cont2_scattering_BBX} (f)], where comparatively deeper bound trimer states are formed \cite{sandoval_radii2D_2016} [see also Fig. \ref{Fig:Spectra} (c)].
In these deep trimer states, the overall size of both species, as captured by the hyperradius $R$ decreases for larger $a_{BB}/a_{BX}$. 
Therefore, the two identical bosons approach each other 
and become strongly correlated.

To address the aforementioned upper bound in the two-body correlations for non-trimer states in BBX and FFX systems, the Jeffreys-Wentzel-Kramers-Brillouin (JWKB) method (see also the review of Ref.~ \cite{friedrich_scattering_2013} and references therein) is employed.
Specifically, the hyperradial part of the three-body wave function [Eq. \eqref{Eq:Hyperradial}] of the atom-dimer states reads 
\begin{equation}
    F^{\rm{JWKB}}_{\nu}(R)=\begin{cases}     
    \frac{1}{\sqrt{p(R)}}\exp \left( -\abs{\int_{R_{\rm{ctp}}}^R p(R')dR'} \right), \, E<U_{\nu}(R) \\
    \frac{2}{\sqrt{p(R)}} \cos \left( \int_{R_{\rm{ctp}}}^R p(R')dR'-\phi \right), \, E>U_{\nu}(R).
    \end{cases}
    \label{Eq:JWKB}
\end{equation}
In the above equation, $\phi=\pi/2+\pi(s_{\nu}-\sqrt{s_{\nu}^2-1/4})$, $p(R)= \sqrt{2\mu \abs{E-U_{\nu}(R)}}$ is the local momentum of a fictitious particle with mass $\mu$ and $R_{\rm{ctp}}$ is the classical turning point, where $E=U_{\nu}(R_{\rm{ctp}})$. Moreover, we focus only on the adiabatic hyperspherical potential curves that support atom-dimer states. In this way, we neglect all the involved non-adiabatic couplings such that we can neatly attribute the upper bound of the contact to type-II atom-dimer states supported by the potentials featuring an atom-dimer threshold. Evidently, the two-body contact [Eq. \eqref{Eq:Cont2}], derived within the JWKB method [Eq. \eqref{Eq:JWKB}] both for BBX and FFX systems [black dashed lines in Figs. \ref{Fig:Cont2_scattering_BBX} and \ref{Fig:Cont2_scattering_FX}], accounts well for the upper bound of $\mathcal{D}_2^{\sigma\sigma'}$ independently of the value of the scattering length.

In order to demonstrate the physical origin of the upper bound in $\mathcal{D}_2^{\sigma\sigma'}$, an approximation for the two-body contact of type-II atom-dimer states is employed in the limit of large inverse inter- and intraspecies scattering lengths $1/a_{\sigma\sigma'}$ [for details see also Appendix \ref{Ap:Upper_bound}]. 
More specifically, and similarly to the approximation employed within the JWKB method, we single out only the adiabatic hyperspherical potential curves supporting atom-dimer states, neglecting the corresponding non-adiabatic couplings. 
Under these considerations, the two-body contact between the $\sigma\, \sigma'$ species ($\sigma\sigma'$=B, F, X) which characterize only atom-dimer eigenstates acquires the following compact form
\begin{equation}
    \mathcal{D}^{\sigma\sigma'}_2 \approx \frac{16 \pi e^{-2\gamma}}{a_{\sigma\sigma'}^2}.  \label{Eq:Upper_bound}\\  
\end{equation} 

\begin{figure*}[t]
    \centering
    \includegraphics[width=1 \textwidth]{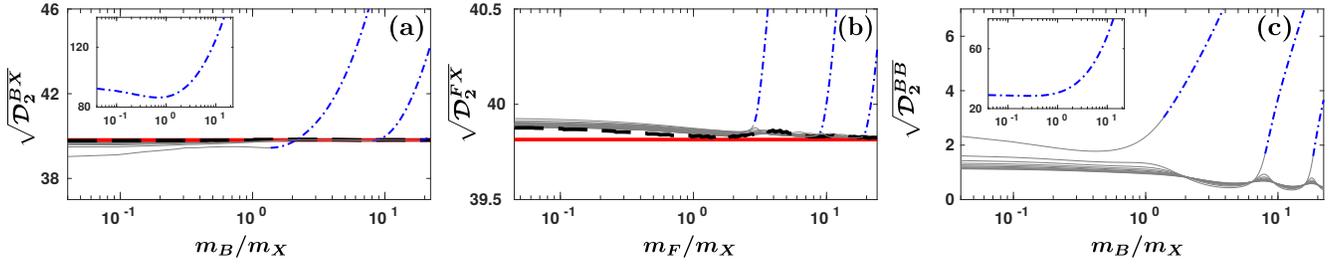}
    \caption{Two-body contact between (a) BX, (b), FX and (c) BB species with respect to the mass ratio, (a)-(b) $m_B/m_X$ and (c) $m_F/m_X$ respectively. The considered scattering lengths are chosen such that (a), (c) $a_{BB}/a_{BX}=10$ and (b) $1/a_{FX}=10$. The inset in panels (a), (c) present $\sqrt{\mathcal{D}_2^{B\sigma}}$ of the ground trimer state for $\sigma=$ X, B respectively. The analytical upper bound [Eq. \eqref{Eq:Upper_bound}] corresponds to the red solid lines and the bound stemming from the JWKB method is showcased in black dashed lines. The two-body contact of (non) trimer states is depicted with (gray solid) blue dash-dotted lines.}
    \label{Fig:Cont2_mass}
\end{figure*}

The prediction of the preceding analytical expression is indeed in good agreement with the upper bound of $\mathcal{D}_2^{BX}$ and $\mathcal{D}_2^{FX}$ within $1-2\%$, except for small scattering lengths $a_{BB}/a_{BX}$ [see the red solid lines in Figs. \ref{Fig:Cont2_scattering_BBX} (a)-(c)] and $1/a_{FX}$ [red solid lines in \cref{Fig:Cont2_scattering_FX} (a)-(c)] respectively. 
Deviations larger than $10\%$ occur up to $a_{BB}/a_{BX}=3$ for LLH and EM, and $a_{BB}/a_{BX}=1.5$ for HHL BBX systems. Likewise, similar deviations are found in the range $1/a_{FX}=[0.36,2]$ for LLH and EM, and $1/a_{FX}=[0.36,1.4]$ for HHL FFX mixtures. 
This discrepancy originates from the fact that the analytical expression [Eq. \eqref{Eq:Upper_bound}] stems from a treatment of the atom-dimer state as a product of the BX (FX) dimer and the third particle. This approach becomes more accurate for large $a_{BB}/a_{BX}$ ($1/a_{FX}$), where the third particle is far away from the strongly bound dimer. 
However, it fails for small $a_{BB}/a_{BX}$ ($1/a_{FX}$), where a product state is not adequate anymore, since the third particle approaches the bound dimer and affects the BX (FX) two-body correlations. In this regime, the JWKB method [Eq. \eqref{Eq:JWKB}] accounts well for the upper bound.
Thus, as suggested by Eq. \eqref{Eq:Upper_bound}, two-body correlations between the BX and FX species depend quadratically on $a_{BB}/a_{BX}$ and $1/a_{FX}$, respectively, in the limit where the  latter two are large. 
This is a manifestation of the universal relation connecting the energy change of an eigenstate with respect to the scattering length and the two-body contact of this state \cite{castin_general_2012,valiente_universal_2011}. 

Interestingly, the analytically obtained upper bound [Eq. \eqref{Eq:Upper_bound}], suggests that $\mathcal{D}_2^{\sigma\sigma'}$ of the atom-dimer states does not depend on the mass ratio of the two identical particles (B,F) with respect to the distinguishable one (X) in the limit where large scattering lengths $a_{BB}/a_{BX}$ and $1/a_{FX}$ are considered. 
To further address this point, the two-body contact versus $\mathcal{M}=\frac{m_{B/F}}{m_X}$ between the BX, BB and FX species is unraveled for large values of the involved scattering lengths i.e., $a_{BB}/a_{BX}, 1/a_{FX}=10$ [Fig. \ref{Fig:Cont2_mass}]. 
Furthermore,  we investigate not only the two-body contact of the atom-dimer states but also of the trimer ones [blue dashed-dotted lines in Fig. \ref{Fig:Cont2_mass}]. 
The latter are naturally included since atom-dimer states convert to trimers in the transition from LLH to HHL of BBX and FFX settings [see also Fig. \ref{Fig:Spectra_mass}].

In the case of trimer states, the two-body contacts increase with $\mathcal{M}$, see the blue dash-dotted lines in \cref{Fig:Cont2_mass}. This behavior is expected since in the HHL scenario the trimer states become deeply bound for both BBX and FFX systems [Fig. \ref{Fig:Spectra} (c), (f)] as $\mathcal{M}$ increases. Moreover, for large mass ratio additional trimer states are formed [Fig. \ref{Fig:Spectra_mass}], whose two-body correlations subsequently shoot up [Fig. \ref{Fig:Cont2_mass}]. 
Indeed, for increasing mass ratio the repulsive wall present at small $R$ [Fig. \ref{Fig:Pots}], recedes to even smaller hyperradii $R$, and so the overall system size of these newly formed trimers decreases, resulting in enhanced two-body correlations. 
Turning to the two-body contact of atom-dimer states between the BX and FX species, depicted with gray solid lines in Figs. \ref{Fig:Cont2_mass} (a), (b), we observe a good agreement between the derived analytical expression [red solid lines in Fig. \ref{Fig:Cont2_mass} (a), (b)] and $\mathcal{D}_2^{BX}$, $\mathcal{D}_2^{FX}$, as well as with the two-body contact obtained via the JWKB method [black dashed lines in Fig. \ref{Fig:Cont2_mass} (a), (b)]. 
Hence, the two-body contact of atom-dimer states in these systems is almost insensitive to a change in the mass ratio, $m_B/m_X$ and $m_F/m_X$.

Moreover, the response of the intraspecies two-body correlations of atom-dimer states in BBX systems as captured by $\mathcal{D}_2^{BB}$ is studied with respect to the mass ratio [depicted with gray solid lines in Fig. \ref{Fig:Cont2_mass} (c)]. Due to the large energy separation of the first and second adiabatic potentials [blue solid and red dash-dotted lines respectively in Fig. \ref{Fig:Pots} (a)-(c)] at this large scattering length ratio $a_{BB}/a_{BX}=10$, we investigate the two-body contact of those eigenstates that lay below the BB+X atom-dimer threshold, exhibited by $U_2(R)$. In order to observe the upper bound in the BB two-body contact, $\sqrt{\mathcal{D}_2^{BB}} \simeq 4$, derived in Eq. \eqref{Eq:Upper_bound}, a larger number of excited eigenstates is required. The intraspecies two-body contact varies mainly in the HHL scenario. At the mass ratio where new trimer states are formed, $\sqrt{\mathcal{D}_2^{BB}}$ of atom-dimer states is enhanced, and subsequently decreases [Fig. \ref{Fig:Cont2_mass} (c)]. This behavior can be attributed to a slight energy attraction and consequent repulsion of the atom-dimer states towards the BX+B atom-dimer threshold as $m_B/m_X$ approaches and further departs from the value where new trimer states are formed. Whenever this slight attraction occurs, and the three-particle system approaches the threshold of trimer state formation, the probability cloud of both species, as captured by $R$, shrinks. Hence, the two identical bosons come closer signaling the increase of $\sqrt{\mathcal{D}_2^{BB}}$. Notice that this pattern emerges also in the interspecies two-body contact, $\sqrt{\mathcal{D}_2^{BX}}$ [Fig. \ref{Fig:Cont2_mass} (a)], being however less pronounced than in the case of the intraspecies one.
Since $a_{BB}=1$, the second adiabatic potential $U_2(R)$ with a BB+X atom-dimer threshold is shallower than the first one which possesses a BX+B threshold, and thus it is more sensitive to the mass ratio. This sensitivity is reflected to the hyperradial part of the wave function, $F_2(R)$, which in turn determines the two-body contact between the identical bosons [see also Eq. \eqref{Eq:Cont2}].

\subsection{Response of the three-body correlations}\label{3b_contact}

\begin{figure*}[t]
    \centering
    \includegraphics[width=1\textwidth]{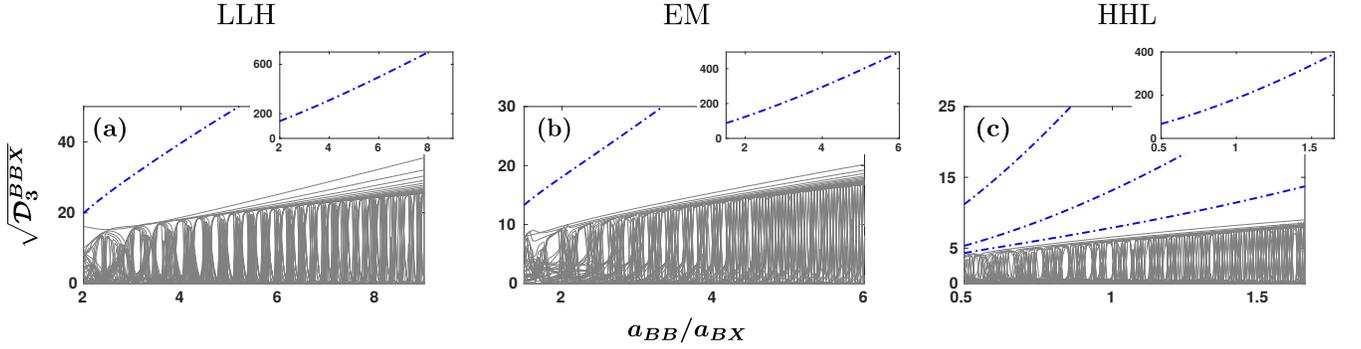}
    \caption{Three-body contact, $\sqrt{\mathcal{D}_3^{BBX}}$ of BBX systems for mass ratio belonging to the (a) LLH ($m_B/m_X=0.04$), (b) EM ($m_B/m_X=1.16$) and (c) HHL ($m_B/m_X=22.16$) class for varying $a_{BB}/a_{BX}$. The three-body contact of (non) trimer states is denoted by (gray solid) blue dash-dotted lines. The insets display the three-body contact of the first trimer state.}
    \label{Fig:Cont3_scattering}
\end{figure*}

As argued above [see Eq.~\eqref{Eq:Asymptotic_expansion_1}], in the asymptotic expansion of the $\sigma$ species reduced one-body density in momentum space there is a contribution related to three-body correlations having an explicit dependence on the single-particle momentum $p_{\sigma}$ [Eq.~\eqref{Eq:Asymptotic_expansion_2}].
Indeed, as it has also been demonstrated in Refs. \cite{Bellotti_contacts_2014,bellotti_dimeffects_2013} treating the three-body problem in momentum space, the next-to-leading order term in the asymptotic expansion of the reduced one-body density in momentum space reads

\begin{equation}
    n^{b}_{\sigma}(\boldsymbol{p}_{\sigma})=\frac{\ln^3p_{\sigma}}{p_{\sigma}^6} \mathcal{D}_3.
    \label{Eq:Cont3}
\end{equation}
In this expression, $\mathcal{D}_3$ is the three-body contact which captures the three-body correlations between all particles that participate in the binary 2D mixture. 
Herein, the three-body contact is derived by linear fitting to $n^{b}_{\sigma}(\boldsymbol{p}_{\sigma})/\ln^3p_{\sigma}$ stemming from the numerical solution of Eq. \eqref{Eq:Asymptotic_expansion_2} for large $p_{\sigma}$.
For the binary mixtures that we consider, the only relevant three-body contact is the one of BBX systems, denoted hereafter by $\mathcal{D}_3^{BBX}$.
For the FFX setting, the three-body correlations are predominantly suppressed due to the Pauli exclusion principle between the identical fermions \cite{Bellotti_contacts_2014}.

Three-body correlations are greatly enhanced when two identical bosons and the third distinguishable particle reside in a trimer state independently of the considered mass ratio [blue dash-dotted lines in Fig. \ref{Fig:Cont3_scattering}]. 
More specifically, in the transition from LLH to the HHL scenario, the first trimer state [insets of Fig. \ref{Fig:Cont3_scattering}] displays a substantially enhanced three-body contact, $\sqrt{\mathcal{D}_3^{BBX}}$. 
This is due to the fact that for heavier identical bosons than the third particle, trimer states become deeply bound as shown in Fig. \ref{Fig:Spectra} (a)-(c).
Hence, all three particles are confined within a small hyperradius which results into large valued three-body contacts. However, in the EM case, the three-body contact of the first trimer state is slightly suppressed compared to the one in the LLH setting [e.g. see the insets in Figs. \ref{Fig:Cont3_scattering} (a) and (b) at $a_{BB}/a_{BX}=2$]. This behavior will be further addressed below arguing on the dependence of $\sqrt{\mathcal{D}_3^{BBX}}$ with respect to the mass ratio. Even though three-body correlations are significantly pronounced for trimer states, the asymptotic expansion of the $\sigma$ species one-body density $n_{\sigma}(\boldsymbol{p}_{\sigma})$ in momentum space [Eq. \eqref{Eq:Asymptotic_expansion_1}] is mainly dominated by the first term attributed to two-body correlations [Eq. \eqref{Eq:Cont2}]. Deviations from the first term occur at $p_{\sigma} \simeq 100$, and especially by considering large scattering length ratios $a_{BB}/a_{BX}>6$, where three-body correlations of the trimer states are more pronounced [see insets of Fig. \ref{Fig:Cont3_scattering}].

Furthermore, three-body correlations of highly excited trap states are greatly reduced, more than two orders of magnitude compared to the three-body contact of trimer states. The large suppression of $\sqrt{\mathcal{D}_3^{BBX}}$ is due to the fact that the system size (captured by the hyperradius $R$) in each of these excited trap states is large. 
As such, the simultaneous collisions of all three particles at small distances become very improbable. In contrast,  atom-dimer states showcase prominent three-body correlations, especially by tuning $a_{BB}/a_{BX}$ to large values. 
In this regime, the atom-dimers [Fig. \ref{Fig:Spectra} (a)] consist of a deeply bound BX dimer accompanied by the second identical bosonic particle. Due to the bosonic symmetry, the BX dimer involves both B atoms, and hence $\sqrt{\mathcal{D}_3^{BBX}}$ increases with the ratio $a_{BB}/a_{BX}$, similarly to the two-body contact. Let us note that in the case of atom-dimer and trap states, the asymptotic expansion of $n_{\sigma}(\boldsymbol{p}_{\sigma})$ for large $p_{\sigma}$ is practically dominated solely by the first term described by Eq. \eqref{Eq:Cont2} being   associated with two-body correlations. 

Similarly to the behavior of the two-body contact, the three-body one of type-I atom-dimer and trap states displays oscillations due to the character change of the latter at the location of the avoided-crossings taking place at specific scattering lengths in the three-body eigenspectrum [see circle in Fig. \ref{Fig:Spectra} (a)]. The lower bound eventually approaches zero for higher lying excited trap eigenstates.  
Recall that an equivalent behavior is observed for the two-body contact $\sqrt{\mathcal{D}_2^{BX}}$ [see Fig. \ref{Fig:Cont2_scattering_BBX} (a)-(c)]. However, in contrast to $\sqrt{\mathcal{D}_2^{BX}}$, the type-II purely atom-dimer states do not provide a well defined upper bound for $\sqrt{\mathcal{D}_3^{BBX}}$. As the ratio $a_{BB}/a_{BX}$ increases, three-body correlations of purely atom-dimer states exhibit a state dependent growth rate [Fig. \ref{Fig:Cont3_scattering}]. The latter is larger for purely atom-dimer states lying close to the BX+B dimer threshold. The three particles residing in these atom-dimer states, are confined within a smaller hyperradius $R$ 
when compared to excited atom-dimers, and as such they feature an enhanced three-body contact. Energetically higher atom-dimer eigenstates are more delocalized, thus possessing a smaller $\sqrt{\mathcal{D}_3^{BBX}}$.

\begin{figure}[t]
    \centering
    \includegraphics[width=0.5 \textwidth]{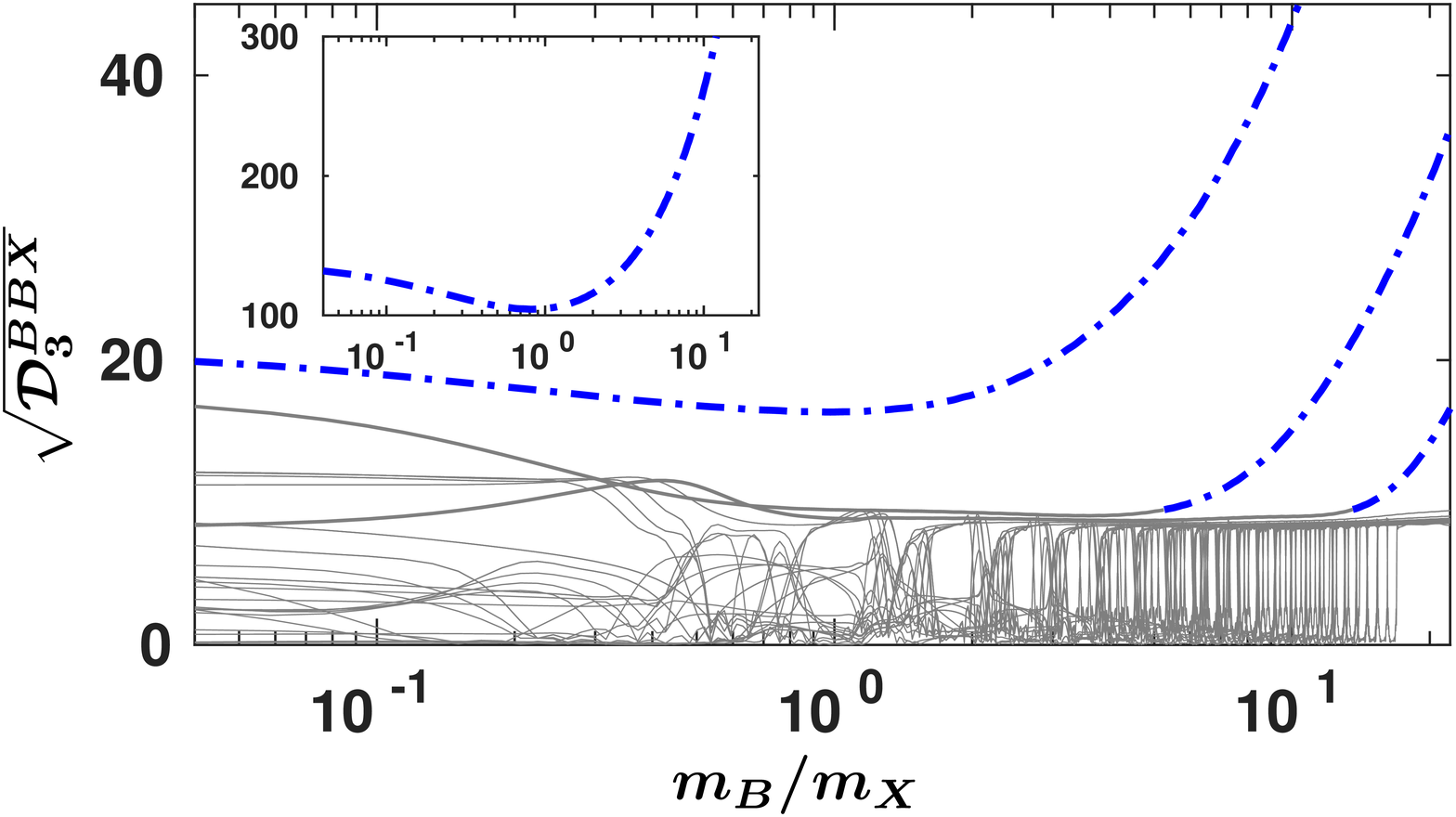}
    \caption{Three-body contact of the BBX system, $\sqrt{\mathcal{D}_3^{BBX}}$, with respect to the mass ratio $m_B/m_X$. The (non) trimer states are denoted with (gray solid) blue dash-dotted lines. The inset displays three-body correlations of the first trimer state. The scattering length ratio reads $a_{BB}/a_{BX}=2$.}
    \label{Fig:Cont3_mass}
\end{figure}

The dependence of $\sqrt{\mathcal{D}_3^{BBX}}$ with respect to the mass ratio of (non) trimer states denoted by blue dash-dotted lines (gray solid lines) is provided in Fig. \ref{Fig:Cont3_mass} exemplarily for $a_{BB}/a_{BX}=2$.
In particular, we observe that an enhancement of three-body correlations takes place in the LLH to the HHL transition for trimer states [blue dash-dotted lines in Fig. \ref{Fig:Cont3_mass}], as discussed previously. 
For sufficiently large $m_B/m_X$, atom-dimer states [denoted by gray lines in Fig. \ref{Fig:Cont3_mass}] change character to trimers [Fig. \ref{Fig:Spectra_mass} (a)], whose three-body correlations subsequently become dominant [see for instance Fig. \ref{Fig:Cont3_mass} at $m_B/m_X \simeq 6$]. This is similar to the enhancement of two-body correlations between both BX and BB species of atom-dimer states when transitioning to trimers as manifested in Fig. \ref{Fig:Cont2_mass} (a) and (c). In particular, the first atom-dimer state, possesses a dominant $\sqrt{\mathcal{D}_3^{BBX}}$, in the LLH regime ($m_B/m_X \simeq 0.04$), similarly to the second trimer state [Fig. \ref{Fig:Cont3_mass}]. Later on, three-body correlations become substantial at the mass ratio where this atom-dimer state transits into a trimer [Fig. \ref{Fig:Cont3_mass} at $m_B/m_X \simeq 6$]. This behavior is caused by an energy shift of both the first atom-dimer and second trimer states towards the BX+B dimer threshold when $m_B/m_X \simeq 0.04$ (LLH), leading to an increased (reduced) three-body contact. 
Notice that the aforementioned energy shift towards the BX+B dimer threshold takes place also for the first trimer state in the LLH to EM transition, resulting in a slight decrease of $\sqrt{\mathcal{D}_3^{BBX}}$ [see the insets of Fig. \ref{Fig:Cont3_mass} and Fig. \ref{Fig:Cont3_scattering} (a)-(b) ]. It is also worth mentioning that $\sqrt{\mathcal{D}_3^{BBX}}$ exhibits oscillations, due to the conversion of trap to type-I atom-dimer states [gray lines in Fig. \ref{Fig:Cont3_mass}] and vice versa nearby the avoided-crossings.

\begin{figure*}[t]
    \centering
    \includegraphics[width=1 \textwidth]{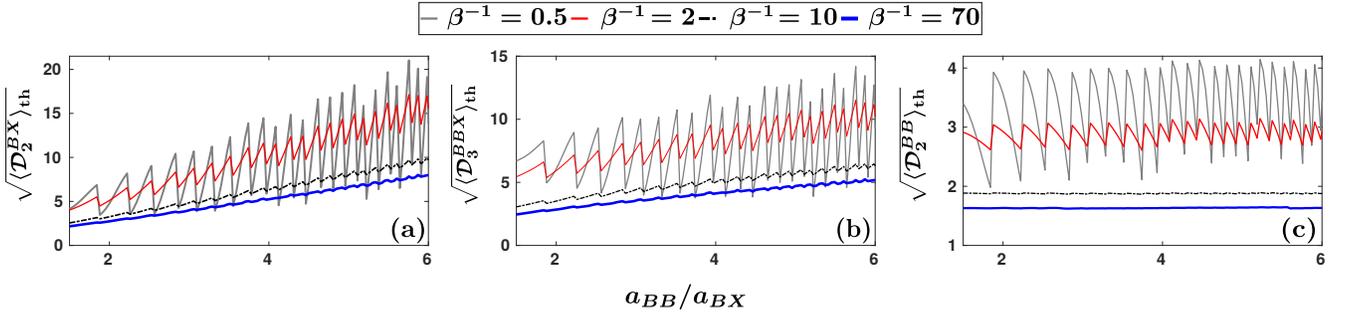}
    \caption{Thermally averaged (a) two-body contact between the BX species, (b) three-body contact and (c) two-body contact between the BB species for an EM BBX system with mass ratio $m_B/m_X=1.16$. The thermal average is performed by employing the Maxwell-Boltzmann distribution for several temperatures $\beta^{-1}=k_B T$ (see legend). Utilizing a radial frequency  $\omega=2\pi \times \, 20 ~\rm{Hz}$, the provided temperatures lie in the range $T=0.48-67.2 ~\rm{nK}$. All observables are expressed in harmonic oscillator units.}
    \label{Fig:Cont_thermal}
\end{figure*}

\subsection{Impact of thermal effects on the correlations}

In the previous subsections, we investigated how two- and three-body correlations depend on the scattering length and mass ratio, as well as, the statistics of the atoms.
Another important issue of immense experimental relevance is the impact of the gas temperature.
Indeed, it has been experimentally and theoretically evinced that temperature effects play a crucial role  on few-body correlation properties of bosonic and two-component fermionic thermal gases \cite{yan_harmonically_2013,sant'ana_scalingTan_2019,zou_planar_Cont_2020,Rakhimov_critical_2021,Hoffman_universality_2015,Capuzzi_finite_2020}. 
For this reason, in the following, the temperature dependence of the two- and three-body contacts with respect to the involved scattering lengths is investigated exemplarily for BBX systems.

The temperature effect, in our system, can be taken into account by simply considering an ensemble of type-I atom-dimer and trap states which are populated according to a Maxwell-Boltzmann distribution \cite{yan_harmonically_2013}.
This means that states with energy $E>0$ are those which can be thermally averaged whereas type-II atom-dimer and trimer states are excluded [see also Sec. \ref{Sec:Spectra} for the classification of the participating states]. This stems from the fact that initially a binary thermal gas consists of unbound atoms with energy larger than zero \cite{Braaten_two-body-loss_2013}. Note that in the limit of a zero trapping frequency we recover the thermally averaged contacts in free space.
Under these considerations, any thermally averaged observable $\braket{\mathcal{O}}_{\rm{th}}$ is given by 

\begin{equation}
    \braket{\mathcal{O}}_{\rm{th}}=\dfrac{\sum_j e^{- \beta E^j}\mathcal{O}^j}{\sum_j e^{-\beta E^j}}~~{\rm{with}}~~\beta^{-1}=k_B T,
    \label{Eq:Thermal}
\end{equation}
where $T$ is the temperature of the gas and $k_B$ is the Boltzmann's constant.
$\mathcal{O}^j$ is the observable associated with the $j$-th eigenstate of our system having an eigenvalue $E^j$, while the summation is performed over eigenstates with $E^j>0$.
Additionally, the range of temperatures that we consider in the following is up to 70 $\rm{nK}$ referring to an experimentally relevant radial trapping frequency $\omega= 2\pi \times \, 20 \: \rm{Hz}$ \cite{jochim_anomalous_2018,jochim_anomalous_2019}. In this temperature regime $s$-wave interactions are adequate for describing thermal effects of few-body correlation observables \cite{zou_planar_Cont_2020}.

The thermally averaged two- and three-body contacts, capturing the imprint  of temperature on the behavior of the two- and three-body correlations respectively, are illustrated in Fig. \ref{Fig:Cont_thermal}. 
We focus on a BBX system with fixed mass ratio $m_B/m_X=1.16$ (EM) and inspect different values of temperature within the interval $T=0.48-67.2~\rm{nK}$. However, we note that mass-imbalanced BBX or FFX systems  display a qualitatively similar behavior. 
For all $\beta^{-1}$ presented in \cref{Fig:Cont_thermal}(a), (b), there is an increasing tendency of two- ($\sqrt{\braket{\mathcal{D}_2^{BX}}_{\rm{th}}}$) and three-body ($\sqrt{\braket{\mathcal{D}_3^{BBX}}_{\rm{th}}}$) correlations between the BX species and the BBX atoms respectively, for a larger scattering length ratio $a_{BB}/a_{BX}$.
This overall behavior stems from the growth observed in the two- and three-body zero temperature contacts [see Figs. \ref{Fig:Cont2_scattering_BBX} (b), \ref{Fig:Cont3_scattering} (b)] associated with type-I atom-dimer states, lying close to the zero energy threshold. 
These eigenstates possess energy larger than zero and thus similarly to the zero temperature scenario they contribute to the growth of the thermally averaged two-body contact.

In particular, $\sqrt{\braket{\mathcal{D}_2^{BX}}_{\rm{th}}}$ features an oscillatory behavior with respect to $a_{BB}/a_{BX}$, whose amplitude decreases as $\beta^{-1}$ becomes larger.
These oscillations originate from the undulations present in the $\sqrt{\mathcal{D}_2^{BX}}$ [see inset of Fig. \ref{Fig:Cont2_scattering_BBX} (a)]. 
They are bounded above by the two-body contact of type-II atom-dimer states and below eventually by zero, in the limit where infinitely many trap states are taken into account. 
In contrast, as the temperature increases, a larger number of eigenstates contributes to the thermal average [Eq. \eqref{Eq:Thermal}] resulting in an $\sqrt{\braket{\mathcal{D}_2^{BX}}_{\rm{th}}}$ free from the oscillatory fringes.
Indeed, in \cref{Fig:Cont_thermal} (a), the $\sqrt{\braket{\mathcal{D}_2^{BX}}_{\rm{th}}}$ at low temperature, i.e. $\beta^{-1}=0.5$ (gray solid line), exhibits prominent oscillations. 
However, as the temperature increases, e.g.  $\beta^{-1}=70$ (blue solid line), more type-I atom-dimer and trap states participate in the thermal average smearing out any interference feature. The same mechanism is responsible for the decay of the oscillation fringes present in $\sqrt{\braket{\mathcal{D}_3^{BBX}}_{\rm{th}}}$ for larger temperatures [Fig. \ref{Fig:Cont_thermal} (b)].

Moreover, we observe that the magnitude of the thermally averaged two-body contact between the BX species shown in \cref{Fig:Cont_thermal} (a) decreases as the temperature of the gas increases \cite{yan_harmonically_2013}.
For example, focusing on $a_{BB}/a_{BX}=6$, $\sqrt{\braket{\mathcal{D}_2^{BX}}_{\rm{th}}}$  is approximately one order of magnitude smaller than the upper bound of $\sqrt{\mathcal{D}_2^{BX}}$ [\cref{Fig:Cont2_scattering_BBX} (b)] at $\beta^{-1}=70$. Similarly, the amount of three-body correlations, quantified by$\sqrt{\braket{\mathcal{D}_3^{BBX}}_{\rm{th}}}$, also becomes suppressed with increasing temperature. In particular, at $\beta^{-1}=70$, the thermally averaged three-body correlations
are reduced by almost a factor of four compared to $\sqrt{\mathcal{D}_3^{BBX}}$ of the first atom-dimer state at $a_{BB}/a_{BX}=6$ [Fig. \ref{Fig:Cont3_scattering} (b)]. Note also that both the thermally averaged two- and three-body contacts at $\beta^{-1}=70$ are suppressed by two orders of magnitude compared to the respective correlation measures of the first trimer state.
This behavior occurs due to the fact that for increasing temperature the likelihood that the three particles occupy trap states becomes larger. These trap states possess fairly small two- and three-body correlations. 
Thus, the thermal average over such states significantly decreases the magnitude of both $\sqrt{\braket{\mathcal{D}_2^{BX}}_{\rm{th}}}$ and $\sqrt{\braket{\mathcal{D}_3^{BBX}}_{\rm{th}}}$.

A similar qualitative behavior is observed for $\sqrt{\braket{\mathcal{D}_2^{BB}}_{\rm{th}}}$, where again two-body correlations are suppressed for increasing $\beta^{-1}$ [\cref{Fig:Cont_thermal} (c)]. In particular, the oscillation amplitude of $\sqrt{\braket{\mathcal{D}_2^{BB}}_{\rm{th}}}$ reduces with increasing temperature ($\beta^{-1}$).
Also, $\sqrt{\braket{\mathcal{D}_2^{BB}}_{\rm{th}}}$ remains almost constant as a function of $a_{BB}/a_{BX}$ for $\beta^{-1}>10$. 
This stems from the fact that the zero-temperature $\mathcal{D}_2^{BB}$ of type-I atom-dimer and trap states does not show an increasing tendency with respect to the tuning of $a_{BB}/a_{BX}$ [see Fig. \ref{Fig:Cont2_scattering_BBX} (d)-(f)], in contrast to the zero-temperature contact of the BX species. Upon increasing the temperature, $\beta^{-1}$, the oscillations of $\mathcal{D}_2^{BB}$ are smeared out [see also the relevant discussion on $\sqrt{\braket{\mathcal{D}_2^{BX}}_{\rm{th}}}$], yielding thus a constant $\sqrt{\braket{\mathcal{D}_2^{BB}}_{\rm{th}}}$. 
Similarly to the case of two-body correlations between the BX species, as the temperature increases the magnitude of the thermally averaged two-body correlations between the identical bosonic particles is further suppressed. In particular, at $\beta^{-1}=70$ [blue solid line in Fig. \ref{Fig:Cont_thermal} (c)], the magnitude of $\sqrt{\braket{\mathcal{D}_2^{BB}}_{\rm{th}}}$ is smaller by a factor of two than the upper bound of two-body correlations between the identical bosonic particles for $a_{BB}/a_{BX}=[1.5,6]$ [Fig. \ref{Fig:Cont2_scattering_BBX} (e)].

\section{Spatial configurations of the three-body states} \label{Sec:dens}

Next, we explore the underlying spatial structure of the few-body binary systems via the corresponding $\sigma$ species one-body reduced density. 
This quantity is a common experimental observable which can be measured by averaging over a sample of different single-shot realizations \cite{Bergschneider_spin_resolved_2018,Anderegg_tweezer_2019,mistakidis2018correlation}, shedding light into the static and dynamical properties of a system \cite{stewart_contact_2010,Katsimiga_bent2D_2017}. 
Within the used descriptive notation [see also Table \ref{Tab:Symmetries} and Appendix \ref{Ap:Asymptotic} for details] the reduced one-body density acquires the form
\begin{eqnarray}
     n_{\sigma}(\boldsymbol{r}_{\sigma})&=& \frac{M}{\pi} \int dR \, d\Omega^{\sigma''\sigma'}\: e^{-G(\boldsymbol{r}_{\sigma},R,\Omega^{\sigma''\sigma'})} \nonumber \\
     &~& \times \abs{\sum_{\nu} F_{\nu} \Phi_{\nu} (R; \Omega^{\sigma''\sigma'})}^2, \label{Eq:dens1}
\end{eqnarray}
where $\sigma' \neq \sigma$, $\sigma''=B/F$ depending on the mixture and $(r_{\sigma},\phi)$ are the polar coordinates of the $\boldsymbol{r}_{\sigma}$ 2D vector. 
Moreover the expression $G(\boldsymbol{r}_{\sigma},R,\Omega^{\sigma''\sigma'})$ reads
\begin{eqnarray}
     G(\boldsymbol{r}_{\sigma},R,\Omega^{\sigma''\sigma'})&=& Mr_{\sigma}^2+(m_{\sigma''}+m_{\sigma'})^2 \frac{R^2\cos^2 \alpha^{\sigma''\sigma'}\mu_{\sigma''\sigma'}}{M \mu} \nonumber \\ &~&+2\frac{(m_{\sigma''}+m_{\sigma'})\sqrt{\mu_{\sigma''\sigma'}}}{\sqrt{\mu}} r_{\sigma} R \cos \alpha^{\sigma'' \sigma'} \nonumber \\
     &~& \times \cos (\theta_2^{\sigma''\sigma'}-\phi).
     \label{Eq:dens2}
\end{eqnarray}
Note that the $\sigma$ species one-body reduced density is normalized to unity.

\begin{figure}[t]
    \centering
    \includegraphics[width=0.5 \textwidth]{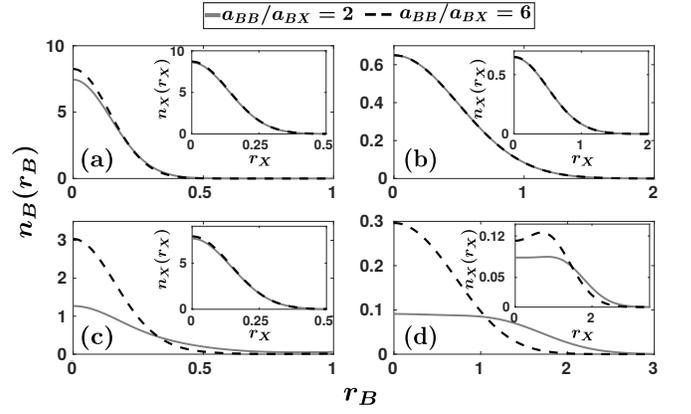}
    \caption{One-body reduced density of (a), (b) the first trimer ($j=1$) and (c), (d) the first atom-dimer states ($j=3$ and $j=5$ respectively) for different scattering length ratios $a_{BB}/a_{BX}$ (see legend). The main panels refer to the density of the B species, $n_B(r_B)$, whereas the insets to the X species, $n_X(r_X)$. The mass ratio of the BBX system is (a), (c) $m_B/m_X=0.04$ (LLH) and (b), (d) $m_B/m_X=22.16$ (HHL).} 
    \label{Fig:Dens_BBX_LLH_HHL}
\end{figure}

Initially, we consider a BBX system with a small mass ratio $m_B/m_X$ i.e. a LLH case. 
Characteristic one-body densities of trimer states are provided in \cref{Fig:Dens_BBX_LLH_HHL} (a) for two representative scattering length ratios $a_{BB}/a_{BX}$.  
Since the X particle is heavier than the identical bosons, it is located close to the trap center while being insensitive to scattering length alterations as can be seen from the Gaussian profile of $n_X(r_X)$ in the inset of Fig. \ref{Fig:Dens_BBX_LLH_HHL} (a). 
The reduced density of the bosonic species closely resembles and encloses the one of the distinguishable particle while it slightly shrinks as $a_{BB}/a_{BX}$ becomes larger, compare $n_B(r_B=0)$ for $a_{BB}/a_{BX}=6$ and  $a_{BB}/a_{BX}=2$ in Fig. \ref{Fig:Dens_BBX_LLH_HHL} (a). This behavior signifies that the light bosons come very close to the heavier distinguishable particle, which is a feature of the trimer state.

Similar density profiles occur for trimer states in the HHL scenario [Fig. \ref{Fig:Dens_BBX_LLH_HHL} (b)].
Evidently, the one-body densities are wider for HHL  [\cref{Fig:Dens_BBX_LLH_HHL} (b)] than LLH [\cref{Fig:Dens_BBX_LLH_HHL} (a)] settings. 
Indeed, as $m_B/m_X$ increases the trapping potential becomes more shallow [see Eq. \eqref{Eq:Pots_adiabatic}] and thus it leads to a larger spatial extent of the one-body density.  
Contrary to the LLH case, here the bosons are placed near the center of the trap due to their heavier mass.
Apart from this difference both $n_B(r_B)$ and $n_X(r_X)$ possess a Gaussian form being almost unaffected by $a_{BB}/a_{BX}$. 
All particles reside close to each other since the system occupies deep trimer states. 

Turning to the first type-II atom-dimer state, see \cref{Fig:Dens_BBX_LLH_HHL} (c) and (d)], we deduce that in contrast to trimer states [Fig. \ref{Fig:Dens_BBX_LLH_HHL} (a), (b)] the corresponding reduced one-body densities are strongly impacted by scattering length variations.
For example, in the case of a LLH BBX system [inset of \cref{Fig:Dens_BBX_LLH_HHL} (c)], $n_X(r_X)$ features a narrow Gaussian distribution which is not altered when tuning $a_{BB}/a_{BX}$ due to the large mass of the X particle. 
However, $n_{B}(r_{B})$ depends strongly on $a_{BB}/a_{BX}$ exhibiting a large spatial extent at $a_{BB}/a_{BX}=2$, whereas at $a_{BB}/a_{BX}=6$ tends to a Gaussian of small width comparable with the one of $n_X(r_X)$ 
[inset of Fig. \ref{Fig:Dens_BBX_LLH_HHL} (c)]. 
This occurs since in this regime ($a_{BB}/a_{BX}=6$) the configuration of the type-II atom-dimer state consists of a strongly bound BX dimer [see also Fig. \ref{Fig:Spectra} (a)] which results into having a boson localized at the trap center and lying close to the X particle.
An analogous behavior of the density takes place for the first type-II atom-dimer state in the HHL case [Fig. \ref{Fig:Dens_BBX_LLH_HHL} (d)]. 
Namely, for increasing scattering length ratio ($a_{BB}/a_{BX}=6$), the density profiles show a narrower spatial configuration.

\begin{figure}[t]
    \centering
    \includegraphics[width=0.5\textwidth]{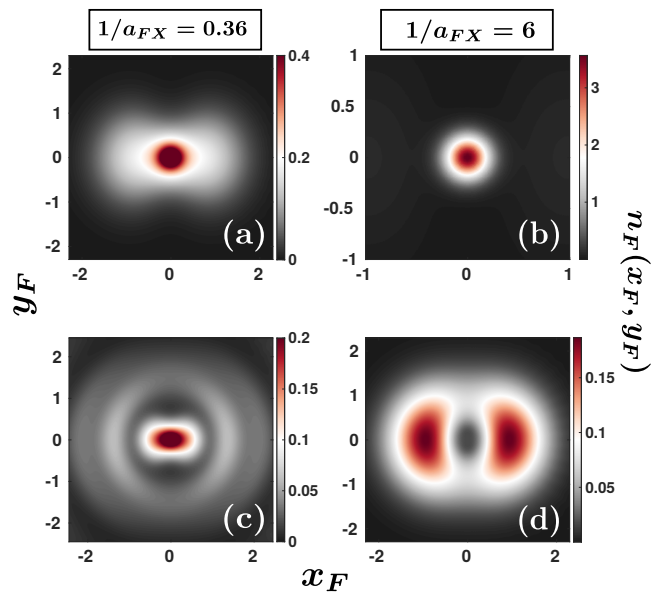}
    \caption{One-body reduced density, $n_F(x_F,y_F)$ of (a), (b) the first atom-dimer ($j=1$) and (c), (d) excited trap states ($j=15$ and $14$ respectively) of a LLH FFX system ($m_F/m_X=0.0451$). The considered interspecies scattering lengths $1/a_{FX}$ read (a), (c) $1/a_{FX}=0.36$ and (b), (d) $1/a_{FX}=6$.}
    \label{Fig:Dens_FFX_A_28_ground_excited}
\end{figure}

The reduced one-body densities of FFX systems presented in \cref{Fig:Dens_FFX_A_28_ground_excited} for small mass ratios ($m_F/m_X=0.0451$, LLH) evince a remarkable angular dependence. 
This is in sharp contrast to the densities of BBX systems which are isotropic and their involved mass and scattering length ratios impact only their radial part.
This difference between the one-body densities of FFX and BBX systems mainly stems from the fact that the total angular momentum and parity of FFX systems is $L^{\pi}=1^-$ whereas for BBX is equal to $L^{\pi}=0^+$. 
Paradigmatic densities of an FFX system occupying the eigenstates of the first atom-dimer [see panels (a) and (b)] and two excited trap states [see panels (c) and (d)] at two different scattering lengths $1/a_{FX}$ are showcased in \cref{Fig:Dens_FFX_A_28_ground_excited}. 
We should note that \cref{Fig:Dens_FFX_A_28_ground_excited} presents only the fermionic density since the one of the X particle features an angular isotropic configuration localized close to the trap center similarly to the structure illustrated in Fig. \ref{Fig:Dens_FFX_A_28_ground_excited} (b). 
This angular isotropy originates from the fact that the X particle interacts with an $s$-wave zero-range pseudopotential with the identical particles and is not constrained by any symmetry as is the case with the identical fermions. 
In contrast, apart from symmetric $s$-wave interactions, particle exchange antisymmetry constraints induce the angular dependence of $n_F(x_F,y_F)$ by introducing a non-zero angular momentum ~\cite{xie_analysis_1997}.

Focusing on the first atom-dimer state [Fig. \ref{Fig:Dens_FFX_A_28_ground_excited} (a), (b)], $n_F(x_F,y_F)$ displays an angular dependent pattern which tends to an isotropic configuration as the interspecies scattering length $1/a_{FX}$ is tuned to a larger value [Fig. \ref{Fig:Dens_FFX_A_28_ground_excited} (b)]. 
In particular, for $1/a_{FX}=0.36$ a small anisotropy is present in the angular direction and $n_{F}(x_{F},y_{F})$ extends to larger distances compared to $1/a_{FX}=6$ [Fig. \ref{Fig:Dens_FFX_A_28_ground_excited} (b)]. 
This is caused by the smaller binding energy of this state compared to the type-II atom-dimer state considered at $1/a_{FX}=6$ [see also Fig. \ref{Fig:Spectra} (d)]. 
Therefore, the densities of type-II atom-dimer states exhibit a configuration where the F particles reside in the vicinity of the trap center, at the location of the X particle, with a larger probability than the respective type-I states [Fig. \ref{Fig:Dens_FFX_A_28_ground_excited} (a)].

The angular deformation of the densities $n_F(x_F,y_F)$ of trap states [Fig. \ref{Fig:Dens_FFX_A_28_ground_excited} (c), (d)], becomes even more pronounced. 
Specifically, for $1/a_{FX}=0.36$ [Fig. \ref{Fig:Dens_FFX_A_28_ground_excited} (c)], the presented eigenstate ($j=15$) is a superposition of an atom-dimer and an excited trap state, since its energy lies close to an avoided-crossing [Fig. \ref{Fig:Spectra} (d)]. 
As can be seen this is directly reflected in the fermionic density which displays a peak close to the trap center, at the location of the distinguishable particle. 
However, away from the peak ($x_F=0$, $y_F=0$) the density shows prominent undulations in the angular direction. Strikingly, by singling out a particular $\nu$ in the summation of Eq. \eqref{Eq:dens1}, one can assign the observed patterns in the reduced densities to specific states that are associated with the $\nu$-th adiabatic hyperspherical potential [Fig. \ref{Fig:Pots} (d)]. 
In this way, these undulations 
are attributed to the specific density patterns building upon states of the higher-lying adiabatic hyperspherical potentials $U_{\nu}(R)$ [Fig. \ref{Fig:Pots} (d)] with $\nu>1$. 
The configuration of the three particles associated with these potentials consist of the X particle being located between the two fermions, which are further separated by the former. Similarly, the lowest adiabatic hyperspherical potential $U_1(R)$, supporting atom-dimer states, is responsible for the density peak close to the trap center [Fig. \ref{Fig:Dens_FFX_A_28_ground_excited} (c)]. 

A different angular pattern appears in the fermionic density of a trap eigenstate ($j=14$) [Fig. \ref{Fig:Dens_FFX_A_28_ground_excited} (d)] for larger interspecies scattering lengths e.g. $1/a_{FX}=6$, whose energy lies away from avoided-crossings. 
Here, the fermions are repelled from the trap center, where the heavy X particle is positioned. 
Since this is a trap state, the attraction between the X particle and the fermions is not strong enough to localize both species at the trap center as is the case for atom-dimer states [Fig. \ref{Fig:Dens_FFX_A_28_ground_excited} (a), (b)]. Thus, the fermions form a shell structure~\cite{papp2008tunable,maity2020parametrically} surrounding the X particle [Fig. \ref{Fig:Dens_FFX_A_28_ground_excited} (d)], a process being reminiscent of the phase separation mechanism emerging in many-body Bose-Fermi mixtures~\cite{lous2018probing,viverit2000zero,mistakidis2019correlated}. Here, this configuration is attributed to adiabatic hyperspherical potentials with $\nu>1$ as can be deduced by focusing on specific $\nu$'s in Eq. \eqref{Eq:dens1}. Notice that this pattern characterizes also other highly excited trap states as well. 

\section{Summary and outlook}  \label{Sec:conclusions}

We have provided insights into the behavior of few-body correlations emerging in arbitrary mass-imbalanced three-body binary mixtures confined in a 2D harmonic trap. In particular, the considered mixtures consist of either two identical bosonic (BB) or fermionic (FF) atoms interacting with a third distinguishable one (X) yielding two distinct physical systems, i.e. BBX and FFX respectively.
Utilizing the hyperspherical formalism we discuss the properties of the individual energy spectra and explicate that they can be classified according to three types of energy hyperradial eigenstates. 
Namely, trap states describing three weakly interacting atoms, a dimer accompanied by a spectator atom (atom-dimers), and trimer configurations. 
The few-body correlation properties, as captured by the two- and three-body contacts, are studied with respect to both the 2D scattering lengths of BBX ($a_{BB}/a_{BX}$) and FFX ($1/a_{FX}$) setups as well as the mass ratio between the species. These correlation measures are investigated for all above mentioned energy eigenstates, distinguishing thus our treatment from previous works in 2D where correlation properties of only the trimer states were considered for three-body binary mixtures in free space \cite{Bellotti_contacts_2014}.

Inspecting the characteristics of two-body intra- (BB) and interspecies (BX, FX) correlations, as encapsulated in the respective two-body contacts, we exemplify a distinction between the aforementioned energy eigenstates. 
Indeed, if the system lies in deep trimer states substantial two-body correlations emerge, which become stronger for increasing scattering lengths (either $a_{BB}/a_{BX}$ or $1/a_{FX}$) or larger mass ratio $m_B/m_X$ and $m_F/m_X$. 
Interestingly, in the case of atom-dimer and trap states the two-body correlations feature an upper and a lower bound 
while exhibiting an oscillatory behavior with respect to the scattering length ratio. 
This response originates from the fact that atom-dimer and trap states change character in the vicinity of avoided-crossings present in the energy spectra.
Qualitatively, these characteristics of the two-body correlations are of universal nature regardless the atomic species or particle exchange symmetry.

In particular, the lower bound is attributed to the presence of highly excited trap states and approaches zero when a larger number of them is taken into account in the two-body contacts. 
On the other hand, the upper bound is associated with pure atom-dimer states. This upper bound is successfully addressed by employing the semi-classical JWKB method, tackling both BBX and FFX systems for all considered scattering length ratios.
For large scattering lengths ($a_{BB}/a_{BX}$ or $1/a_{FX}$) an analytical expression for this upper bound is derived, stemming from the approximation of the atom-dimer wave function as a product state describing a deep dimer and the remaining trapped particle. 
Interestingly, the values of the upper bound depends solely on the considered scattering lengths $a_{BB}/a_{BX}$ or $1/a_{FX}$. 
By comparing with the semi-classical JWKB formula, it is possible to infer the effect of the third particle to the two-body correlations of the dimer, especially at small scattering lengths $a_{BB}/a_{BX}$ or $1/a_{FX}$.
Moreover, the two-body contact of atom-dimer states with respect to the mass ratio for large fixed $a_{BB}/a_{BX}$ or $1/a_{FX}$, shows a saturation tendency towards the analytically predicted value.

Turning to BBX mixtures a similar oscillatory behavior of three-body correlations occurs as a function of the scattering length for atom-dimer and trap states as manifested in the three-body contact.
The corresponding lower bound is caused by the existence of energetically higher-lying trap states. 
Antithetically to two-body correlations, the three-body contact of the atom-dimer states lacks an upper bound, exhibiting a state-dependent growth rate. Three-body correlations are more enhanced for atom-dimer states residing close to the trimer formation threshold, compared to states lying further away from it. 
Proceeding one step further, the impact of thermal effects on the two- and three-body correlations of the 2D binary mixtures is also investigated.
Concretely, for thermal gases we observe that a thermally averaged two- or three-body contact possesses an oscillatory pattern with a reduced peak-to-peak amplitude as the temperature increases, whereas their overall magnitude is also suppressed. This behavior is attributed to a superposition of highly excited trap and atom-dimer states weighted according to the Boltzmann distribution, which destroys the oscillatory patterns present for zero temperature.

To comprehend the spatially resolved structure of the species in the different eigenstates we employ the respective reduced one-body densities, an observable that has not been extensively studied in 2D three-body binary systems \cite{sandoval_radii2D_2016}. 
For trimer and atom-dimer states of BBX systems the reduced density displays an isotropic configuration in the angular direction, with a spatial extent characterized by the binding energy of the state. 
On the other hand, for FFX systems patterns with an angular dependence appear in the fermionic reduced density due to the non-zero total angular momentum of the system, which is particularly more prominent for excited trap states. 
However, the density of the distinguishable X particle features an isotropic configuration in the angular direction, since it is not constrained by any symmetry contrary to the case of the fermionic particles and similarly to 
the distinguishable particle in 
BBX systems. 

Concluding, there are many interesting future perspectives that are worth being studied. 
For instance, the investigation of the dynamical formation of trimer and atom-dimer states~\cite{Incao_dynamics_2018} and in particular the interplay and transfer efficiency of the involved two- and three-body correlations~\cite{Colussi_dynamics_2018} by e.g. applying interaction quenches or time-dependent pulses will yield insight into the early-time dynamics of Bose and Fermi gases in 2D. 
Another aspect regards the inclusion of finite-range corrections~\cite{Incao_recomb2D_2015} that would possibly alter the upper bound of the two-body contact for atom-dimer states in the regime of large inverse scattering lengths. 
Indeed, the adiabatic potential curves can exhibit higher-order corrections for finite-range two-body potentials as discussed in Ref.~\cite{Volosniev_Borromean_2014}. 
For example, in the case of the EM FFX system, finite-range effects are important in the region of $1/a_{FX} \gtrsim 55$ for a trapping frequency $\omega=2\pi \times 20 \, \rm{Hz}$, and in the interval $1/a_{FX} \gtrsim 8$ for $\omega=2\pi \times 1 \, \rm{kHz}$ ~\cite{Incao_recomb2D_2015}.
An additional possibility for future studies concerns the stationary properties of fermionic mixtures featuring $p$-wave interactions~\cite{kanjilal_coupled-channel_2006,Volosniev_Borromean_2014,nishida_super_2013} which will permit the exploration of unitary Fermi  gases from a few-body perspective. 
\begin{acknowledgments} 
The authors thank A. G. Volosniev for insightful comments. G. B. kindly acknowledges financial support by the State Graduate Funding Program Scholarships (Hmb-NFG). 
S. I. M. gratefully acknowledges financial support in the framework of  the Lenz-Ising Award of the University of Hamburg. This work is supported by the Cluster of Excellence "CUI: Advanced Imaging of Matter" of the Deutsche Forschungsgemeinschaft (DFG)-EXC 2056-project ID 390715994.
\end{acknowledgments}

\appendix

\section{Boundary condition of the hyperangular part in the hyperspherical formalism}   \label{Ap:Hyperangles}

The boundary condition whenever two particles collide 
is expressed within the hyperspherical formalism.
Since the interparticle interaction is modelled by 
a delta pseudopotential [Eq. \eqref{Eq:Pseudpot}] the hyperangular wave functions satisfying Eq. \eqref{Eq:Hyperangular} can be written in a closed analytical form whenever the particles $i$ and $j$ collide \cite{rittenhouse_greens_2010}, i.e., $\rho_1^{(k)}\to 0$,

\begin{equation}
    \lim_{\rho_1^{(k)}\to 0} \Phi_{\nu}(R;\Omega)=\sum_{l=\pm L} C_{\nu,l}^{(k)}(R)Y_l(\theta_2^{(k)}) Y_0(\theta_1^{(k)}) \ln\left( \frac{d_k \rho_1^{(k)}}{a^{(k)}}\right). \label{Ap:boundary}
\end{equation}

The above equation can be rewritten in the following form,
\begin{eqnarray}\label{bound2}
\sum_{l=\pm L} Y_l(\theta_2^{(k')}) &~&C_{\nu,l}^{(k')} =-\lim_{\rho_1^{(k')}\rightarrow 0} \frac{1}{\ln(A\Lambda a^{(k')})} \nonumber \\ &~& \times \left[1-\ln(A\Lambda d_{k'} \rho_1^{(k')})\rho_1^{(k')}\frac{\partial}{\partial \rho_1^{(k')}} \right] \Phi_{\nu}(R;\Omega), \nonumber \\
\end{eqnarray}
an expression which will ultimately determine the eigenvalues $s_{\nu}$ and the $C_{\nu}^{(k)}$ coefficients.

\section{Asymptotic expansion of the reduced one-body density} \label{Ap:Asymptotic}

The expansion of the reduced one-body density at large single-particle momenta is derived first in a general form in the lab frame. Subsequently, a coordinate transformation is employed so that this asymptotic expansion is expressed within the hyperspherical formalism. 

\subsection{Reduced one-body density in the lab frame}

The reduced one-body density of the $\sigma=$ B,F or X species containing $N_{\sigma}$ particles reads in momentum space
\begin{eqnarray}
   n_{\sigma}(\boldsymbol{p}_{\sigma})&=&\frac{1}{N_{\sigma}}\sum_{i=1}^{N_{\sigma}}\int \prod_{j\neq i} d\boldsymbol{r}_j \, \abs{\tilde{\Psi}_{\sigma}(\boldsymbol{p}_{\sigma},\boldsymbol{r}_{j \neq i})}^2,  \label{Ap:density}
   \end{eqnarray}
with $\tilde{\Psi}_{\sigma}(\boldsymbol{p}_{\sigma},\boldsymbol{r}_{j \neq i})$ being
\begin{eqnarray}
\tilde{\Psi}_{\sigma}(\boldsymbol{p}_{\sigma},\boldsymbol{r}_{j \neq i})&= & \int d\boldsymbol{r}_i \, e^{-i \boldsymbol{p}_{\sigma}\cdot \boldsymbol{r}_i} \Psi_{\textrm{tot}}(\boldsymbol{r}_i,\boldsymbol{r}_j,\boldsymbol{r}_k).
\end{eqnarray}
In the last expression $p_{\sigma}$ is the single-particle momentum of the $\sigma$-species, and $\Psi_{\textrm{tot}}$ is the total wave function in the lab frame, including the center-of-mass contribution. Similarly, the reduced one-body density in position space reads

\begin{equation}
    n_{\sigma}(\boldsymbol{r}_{\sigma})=\int d\boldsymbol{r}_j d\boldsymbol{r}_k \, \abs{\Psi_{\textrm{tot}}(\boldsymbol{r}_{\sigma},\boldsymbol{r}_j,\boldsymbol{r}_k)}^2\label{Eq:Ap:Position}.
\end{equation}
$n_{\sigma}(\boldsymbol{p}_{\sigma})$ can be decomposed into a part where the $i$-th and $j$-th particles approach each other while $j\neq i$ and another part where the integration is performed in the remaining space \cite{dunjko_correlation_2003,patu_universal_2017}. In the first part, the following 2D boundary condition is employed \cite{castin_general_2012}

\begin{equation}
    \Psi_{\textrm{tot}}(\boldsymbol{r}_i,\boldsymbol{r}_j,\boldsymbol{r}_k)\stackrel{\boldsymbol{r}_i \simeq \boldsymbol{r}_j}{\longrightarrow} \ln \left(\frac{r_{ij}}{a_{ij}} \right) A_{ij}(\boldsymbol{c}_{ij},\boldsymbol{r}_{k \neq i,j}),
    \label{Ap:boundary_2}
\end{equation}
where $c_{ij}=\frac{m_i \boldsymbol{r}_i+m_j\boldsymbol{r}_j}{m_i+m_j}$ is the center-of-mass of the $i$-th and $j$-th particles, $\boldsymbol{r}_{ij}=\boldsymbol{r}_i-\boldsymbol{r}_j$ denotes their relative position, $A_{ij}(\boldsymbol{c}_{ij},\boldsymbol{r}_{k \neq i,j})$ is a regular function and $a_{ij}$ signifies the scattering length corresponding to the interaction of the $i-j$ pair. Thus, the reduced one-body density in momentum space acquires the following asymptotic expansion,
\begin{equation}
n_{\sigma}(\boldsymbol{p}_{\sigma}) \approx n^a_{\sigma}(\boldsymbol{p}_{\sigma})+  n^b_{\sigma}(\boldsymbol{p}_{\sigma}),
\end{equation}
which is valid for $p_{\sigma}$ larger than all the momentum scales provided by the scattering lengths $a_{ij}^{-1}$ between the $i-j$, $j\neq i$, particle pairs. The two terms read explicitly
\begin{eqnarray}
   n^a_{\sigma}(\boldsymbol{p}_{\sigma}) =  \frac{4\pi^2}{N_{\sigma} p_{\sigma}^4}\sum_{i=1}^{N_{\sigma}} \sum_{k \neq i} \int \prod_{j \neq i} d\boldsymbol{r}_j  \abs{A_{ik}(\boldsymbol{c}_{ik},\boldsymbol{r}_{j \neq i,k})}^2 \label{Ap:Asymptotic_expansion_1} 
   \end{eqnarray}
   and
 \begin{eqnarray}
  n^{b}_{\sigma}(\boldsymbol{p}_{\sigma}) & =&  \frac{4\pi^2}{N_{\sigma} p_{\sigma}^4} \sum_{i=1}^{N_{\sigma}} \sum_{\substack{k,j \\k\neq j \neq i}} \int \prod_{l \neq i} d\boldsymbol{r}_l \, \exp\left[-i\boldsymbol{p}_{\sigma}\cdot (\boldsymbol{r}_k-\boldsymbol{r}_{j})\right] \nonumber \\
   & & \times A_{ik}(\boldsymbol{c}_{ik},\boldsymbol{r}_{j \neq i,k})A^*_{ij}(\boldsymbol{c}_{ij},\boldsymbol{r}_{k \neq i,j}).
   \label{Ap:Asymptotic_expansion_2}
\end{eqnarray}

\subsection{Transformation to the body-frame}
In order to transform Eqs. \eqref{Ap:Asymptotic_expansion_1}, \eqref{Ap:Asymptotic_expansion_2} to the body-frame, the following coordinate transformation is employed ~\cite{colussi_contacts_2019}

\begin{eqnarray}
     \int d\boldsymbol{r}_jd\boldsymbol{r}_k  &=&\frac{1}{d_j^2}  \int d\boldsymbol{r}_{\textrm{CM}}d\boldsymbol{\rho}_2^{(j)} \nonumber \\
     &=& \frac{1}{d_j^2} \int d\boldsymbol{r}_{\textrm{CM}} \, d\theta_2^{(j)} dR \, R,
     \label{Ap:Coord_trans_1}
\end{eqnarray}
where $\boldsymbol{\rho}_2^{(j)}=d_k(\boldsymbol{r}_k-\boldsymbol{r}_j)$ and $\boldsymbol{r}_{\textrm{CM}}$ is the center-of-mass of the three particles. The norm of the second Jacobi vector in Eq.~\eqref{Ap:Coord_trans_1}, $\rho_2^{(j)}$ is substituted by the hyperradius $R$, since the first Jacobi vector vanishes due to the boundary condition [see also Appendix \ref{Ap:Hyperangles}].

The boundary condition Eq. \eqref{Ap:boundary_2} can be also expressed in the hyperspherical formalism, making use of the descriptive notation $\sigma\sigma'$, denoting the species B,F or X [see also Table \ref{Tab:Symmetries}]. Thus, whenever a $\sigma$ species particle collides with a $\sigma'$ species one ($\alpha^{\sigma\sigma'}\to 0$)
\begin{eqnarray}
    \Psi_{\rm{CM}}(\boldsymbol{r}_{\rm{CM}})&~& \Psi(R;\Omega)\stackrel{\alpha^{\sigma\sigma'}\to 0}{\longrightarrow}\Psi_{\textrm{CM}}(\boldsymbol{r}_{\textrm{CM}})\ln \left(  \frac{\sqrt{\mu} R\alpha^{\sigma\sigma'}}{\sqrt{\mu_{\sigma\sigma'}}a_{\sigma\sigma'}} \right) \nonumber \\ &~& \times \sum_{\nu}\frac{F_{\nu}(R)}{R^{3/2}} \sum_{l=\pm L} C_{\nu,l}^{\sigma\sigma'}(R) Y_l(\theta_2^{\sigma\sigma'}) Y_0(\theta_1^{\sigma\sigma'}) . \nonumber \\
\end{eqnarray}
Here, $\mu_{\sigma\sigma'}$ and $a_{\sigma\sigma'}$ denote the two-body reduced mass and scattering length respectively between the $\sigma\sigma'$ species.
By integrating the center-of-mass, the first term, $n^a_{\sigma}(\boldsymbol{p}_{\sigma})$ is expressed as follows
\begin{eqnarray}
n^a_{\sigma}(\boldsymbol{p}_{\sigma})& =& \frac{4\pi}{\mu N_{\sigma} p_{\sigma}^4}\sum_{\sigma'}\mu_{\sigma\sigma'} \int_0^{\infty} \frac{dR}{R^2} \abs{\sum_{\nu} F_{\nu}(R) \sum_{l=\pm L} C_{\nu,l}^{\sigma\sigma'}(R)}^2 \nonumber \\
& = & \frac{1}{N_{\sigma} p_{\sigma}^4} \sum_{\sigma'} (1+\delta_{\sigma\sigma'}) \mathcal{D}^{\sigma\sigma'}_2 , 
\end{eqnarray}
where $\mathcal{D}^{\sigma\sigma'}_2$ is the two-body contact between the species $\sigma\sigma'$.
Similarly, the second term $n^{b}_{\sigma}(\boldsymbol{p}_{\sigma})$ yields,

\begin{eqnarray}
     n^{b}_{\sigma}(\boldsymbol{p}_{\sigma})&=& \frac{4\pi}{N_{\sigma}p_{\sigma}^4} \sum_{\sigma'}\frac{\mu_{\sigma\sigma'}}{\mu}  \int_0^{\infty} \frac{dR}{R^2} \,\Bigg\{ J_0\left[\frac{p_{\sigma} R \sqrt{\mu_{\sigma\sigma'}}}{\sqrt{\mu}} \right](-1)^L   \nonumber  \\
    &~& +J_{2L}\left[\frac{p_{\sigma} R \sqrt{\mu_{\sigma\sigma'}}}{\sqrt{\mu}} \right](1-\delta_{0,L}) \Bigg \} \nonumber \\
    &~& \times \sum_{\sigma'' \neq \sigma'} \sum_{l= \pm L} \left(\sum_{\nu} F_{\nu}(R) C_{\nu,l}^{\sigma\sigma'}(R) \right) \nonumber \\
    &~& \times \left(\sum_{\nu'} F_{\nu'}(R) C_{\nu',l}^{\sigma' \sigma''}(R) \right)^*
\end{eqnarray}
where $J_{\nu}(\cdot)$ is the $\nu$-th Bessel function of the first kind, and $L$ is the total angular momentum of the system. 

 Regarding the reduced one-body density in position space, by employing the transformation from the lab to the body-frame, $\int \prod_{j \neq i} d\boldsymbol{r}_j=\int d\boldsymbol{\rho}_1^{(i)}d\boldsymbol{\rho}_2^{(i)}$ and the descriptive notation, one gets

\begin{eqnarray}
     n_{\sigma}(\boldsymbol{r}_{\sigma})&=& \frac{M}{\pi} \int dR \, d\Omega^{\sigma''\sigma'}\: e^{-G(\boldsymbol{r}_{\sigma},R,\Omega^{\sigma''\sigma'})} \nonumber \\
     &~& \times \abs{\sum_{\nu} F_{\nu} \Phi_{\nu} (R; \Omega^{\sigma''\sigma'})}^2,
     \label{Ap:dens}
\end{eqnarray}
where
\begin{eqnarray}
     G(\boldsymbol{r}_{\sigma},R,\Omega^{\sigma''\sigma'})&=& Mr_{\sigma}^2+(m_{\sigma''}+m_{\sigma'})^2 \frac{R^2\cos^2 \alpha^{\sigma''\sigma'}\mu_{\sigma''\sigma'}}{M \mu} \nonumber \\ &~&+2\frac{(m_{\sigma''}+m_{\sigma'})\sqrt{\mu_{\sigma''\sigma'}}}{\sqrt{\mu}} r_{\sigma} R \cos \alpha^{\sigma'' \sigma'} \nonumber \\
     &~& \times \cos (\theta_2^{\sigma''\sigma'}-\phi).
\end{eqnarray}
In the above expressions, $\sigma' \neq \sigma$, $\sigma''=B/F$ depending on the mixture and $(r_{\sigma},\phi)$ are the polar coordinates of the $\boldsymbol{r}_{\sigma}$ 2D vector.

\section{Derivation of the upper bound of the two-body contact} \label{Ap:Upper_bound}

In the limit of large inter- and intraspecies scattering lengths $1/a_{\sigma\sigma'}$, an approximate analytical form of the corresponding two-body contact of atom-dimer states is derived.
The adiabatic potentials $U_{\nu}(R)$ with $\nu=1$ ($\nu=1, \, 2$) at large hyperradius $R$ and in the absence of a trap asymptote to an atom-dimer threshold in the case of FFX (BBX) systems. This behavior is reflected to the eigenvalues $s_{\nu}(R)$, which obey the following relations \cite{kartavtsev_universal_2006,Incao_recomb2D_2015},

\begin{eqnarray}
     s_1(R)&  \stackrel{R> R_0}{\longrightarrow}&  i\frac{2e^{-\gamma} R\sqrt{1+\mathcal{M}}}{a_{\sigma X} \sqrt[4]{2\mathcal{M}+1}} \\
	 s_2(R) & \stackrel{R > R_0}{\longrightarrow} & i\frac{2\sqrt{2}e^{-\gamma} R}{a_{BB} \sqrt[4]{2\mathcal{M}+1}},
\end{eqnarray}
where $\mathcal{M}=\frac{m_{F/B}}{m_X}$, $\sigma=B/F$ and $i$ is the imaginary unit. The value of $R_0$ is proportional to the scattering lengths, and so for large $1/a_{FX}$, $1/a_{BX}$, and $1/a_{BB}$, the parameter $R_0$ becomes small. In these regimes, the two-body contact between $\sigma\sigma'$ species can be decomposed into two parts
\begin{eqnarray}
    \mathcal{D}^{\sigma\sigma'}_2&=&\frac{2\pi \mu_{\sigma\sigma'}(2-\delta_{\sigma\sigma'})}{\mu} \int_0^{R_0} \frac{dR}{R^2} \abs{ F_{\nu}(R) \sum_{l=\pm L} C_{\nu,l}^{\sigma\sigma'}(R)}^2 \nonumber \\  & &+\frac{2\pi \mu_{\sigma\sigma'}(2-\delta_{\sigma\sigma'})}{\mu} \int_{R_0}^{\infty} \frac{dR}{R^2} \abs{ F_{\nu}(R) \sum_{l= \pm L} C_{\nu,l}^{\sigma\sigma'}(R)}^2. \nonumber \\
\end{eqnarray}
In the above expression we take into account only the first potential ($\nu=1$) for FFX or the two lowest potentials ($\nu=1,2$) for BBX systems, which support atom-dimer states and neglect all the other coupling elements with the remaining adiabatic potentials.

The $C_{\nu,l}^{\sigma\sigma'}$ coefficients satisfy a semi-analytical expression \cite{kartavtsev_universal_2006,kartavtsev_universal_2007,rittenhouse_greens_2010}, valid in the case of a zero-range pseudopotential, relating the coefficients with the derivatives of the eigenvalues $s_{\nu}$. By employing these expressions we end up with the following relations in the limit where $1/a_{FX}, 1/a_{BX}, 1/a_{BB} \gg 1$,

\begin{eqnarray}
     \mathcal{D}^{\sigma X}_2&=&\frac{4\pi \sqrt{2\mathcal{M}+1}}{1+\mathcal{M}} \int_0^{R_0} \frac{dR}{R^2} \abs{ F_1(R) \sum_{l= \pm L} C_{1,l}^{\sigma X}(R)}^2 \nonumber \\ & &+\frac{16\pi e^{-2\gamma}}{a_{\sigma X}^2}\int_{R_0}^{\infty} dR \, \abs{F_1(R)}^2 \nonumber \\ & & \approx \frac{16\pi e^{-2\gamma}}{a_{\sigma X}^2}.
     \label{Ap:Upper_FFX} \\
    \mathcal{D}^{BB}_2&=&\frac{2 \pi \sqrt{2\mathcal{M}+1}}{2} \int_0^{R_0} \frac{dR}{R^2} \abs{ F_2(R) \sum_{l= \pm L} C_{2,l}^{BB}(R)}^2 \nonumber \\ & &+\frac{16\pi e^{-2\gamma}}{a_{BB}^2}\int_{R_0}^{\infty} dR \, \abs{F_2(R)}^2 \nonumber \\ & & \approx \frac{16\pi e^{-2\gamma}}{a_{BB}^2}, \label{Ap:Upper_BB}
\end{eqnarray}
where $\sigma=B/F$. In the last steps of Eqs. \eqref{Ap:Upper_FFX}, \eqref{Ap:Upper_BB}, we have kept only the dominant second term. The second integral is approximated by unity since $R_0$ is small. 
Hence, this yields the normalization condition for the hyperradial part $F_{\nu}(R)$, where $\nu=1,2$.

\bibliography{references.bib, zot_biblio.bib}

\end{document}